\documentclass[
]{ametsocV6.1}

\usepackage[
    final
    authormarkup=none
]{changes}

\usepackage[
    inner=1in,      
    outer=1in,      
    top=1.3in,        
    bottom=1in,     
    paper=letterpaper
]{geometry}
\usepackage{siunitx}
\usepackage{color}

\DeclareSIUnit{\y}{y}

\definecolor{light}{gray}{0.50}
\definecolor{heavy}{gray}{0.35}
\definecolor{black}{gray}{0.0}
\definecolor{dgreen}{rgb}{0.0,0.5,0}
\definecolor{dred}{rgb}{0.9959,0,0}
\definecolor{green}{rgb}{0.0,0.99599,0.0}
\definecolor{purple}{rgb}{0.6,0.0,0.4}

\newcommand{\reviewcolor}{black}

\definechangesauthor[name=Georg Sebastian, color=\reviewcolor]{GSV}
\definechangesauthor[name=Young-Ha, color=\reviewcolor]{YHK}
\definechangesauthor[name=Ulrich, color=\reviewcolor]{UA}
\definechangesauthor[name=Gergely, color=\reviewcolor]{GB}
\definechangesauthor[name=Günther, color=\reviewcolor]{GZ}

\title{MS-GWaM: A 3-dimensional transient gravity wave parametrization for atmospheric models}

\authors{
    Georg S. Voelker\aff{a}\correspondingauthor{Georg S. Voelker, voelker@iau.uni-frankfurt.de},
    Gergely B\"ol\"oni\aff{b}, 
    Young-Ha Kim\aff{a},
    G\"unther Z\"angl\aff{b},\\
    and Ulrich Achatz\aff{a} 
}

\affiliation{
    \aff{a}{Institute for Atmosphere and Environment, Goethe University Frankfurt, Frankfurt, Germany}\\
    \aff{b}{Deutscher Wetterdienst, Offenbach am Main, Germany}
}

\abstract{
    Parametrizations for internal gravity waves in atmospheric models are traditionally subject to a number of simplifications. Most notably, they rely on both neglecting wave propagation and advection in the horizontal direction (single-column assumption) and an instantaneous balance in the vertical direction (steady-state assumption). While these simplifications are well justified to cover some essential dynamic effects and keep the computational effort small it has been shown that both mechanisms are potentially significant. In particular, the recently introduced Multiscale Gravity Wave Model (MS-GWaM) successfully applied ray tracing methods in a novel type of transient but columnar internal gravity wave parameterization (MS-GWaM-1D). We extend this concept to a three-dimensional version of the parameterization (MS-GWaM-3D) to simulate subgrid-scale non-orographic internal gravity waves. The resulting global wave model---implemented into the weather-forecast and climate code ICON---contains three-dimensional transient propagation with accurate flux calculations, a latitude-dependent background source, and convectively generated waves. MS-GWaM-3D helps reproducing expected temperature and wind patterns in the mesopause region in the climatological zonal mean state and thus proves a viable IGW parameterization. Analyzing the global wave action budget, we find that horizontal wave propagation is as important as vertical wave propagation. The corresponding wave refraction includes previously missing but well-known effects such as wave refraction into the polar jet streams. On a global scale, three-dimensional wave refraction leads to a horizontal flow-dependent redistribution of waves such that the structures of the zonal mean wave drag and consequently the zonal mean winds are modified.
}

\begin{document}

\noindent \textbf{This Work has been submitted to the Journal of Atmospheric Sciences. Copyright in this Work may be transferred without further notice.}

\maketitle



\renewcommand{\v}[1]{\ensuremath{\boldsymbol{#1}}}
\newcommand{\vq}[1]{\v{#1}}
\newcommand{\nab}[1][]{\ensuremath{\nabla_{#1}}}
\newcommand{\nabh}[1][]{\ensuremath{\nabla_{#1\mbox{\scriptsize h}}}}
\newcommand{\nabhh}{\ensuremath{\hat{\nabla}_{\mbox{\scriptsize h}}}}

\newcommand{\vex}{\ensuremath{\v{e}_x}}
\newcommand{\vey}{\ensuremath{\v{e}_y}}
\newcommand{\vez}{\ensuremath{\v{e}_z}}
\newcommand{\vezh}{\ensuremath{\v{e}_{\hat{z}}}}

\newcommand{\dt}{\ensuremath{\partial_t}}
\newcommand{\dth}{\ensuremath{\partial_{\hat{t}}}}
\newcommand{\dT}[1][]{\ensuremath{\partial_{T_{#1}}}}
\newcommand{\dX}[1][]{\ensuremath{\partial_{X_{#1}}}}
\newcommand{\dY}[1][]{\ensuremath{\partial_{Y_{#1}}}}
\newcommand{\dZ}[1][]{\ensuremath{\partial_{Z_{#1}}}}
\newcommand{\dtau}[1][]{\ensuremath{\partial_{\tau_{#1}}}}
\newcommand{\ds}[1][]{\ensuremath{\partial_{s_{#1}}}}
\renewcommand{\dm}{\ensuremath{\partial_{m}}}
\newcommand{\dk}[1][]{\ensuremath{\partial_{k_{#1}}}}

\newcommand{\dxx}{\ensuremath{\frac{\partial^2}{\partial x^2}}}
\newcommand{\dyy}{\ensuremath{\frac{\partial^2}{\partial y^2}}}
\newcommand{\dzz}{\ensuremath{\frac{\partial^2}{\partial z^2}}}
\newcommand{\dx}{\ensuremath{\partial_x}}
\newcommand{\dy}{\ensuremath{\partial_y}}
\newcommand{\dz}{\ensuremath{\partial_z}}
\newcommand{\dzh}{\ensuremath{\partial_{\hat{z}}}}
\newcommand{\dlambda}{\ensuremath{\partial_\lambda}}
\newcommand{\dphi}{\ensuremath{\partial_\phi}}
\newcommand{\dr}{\ensuremath{\partial_r}}
\newcommand{\dklambda}{\ensuremath{\partial_{k_\lambda}}}
\newcommand{\dkphi}{\ensuremath{\partial_{k_\phi}}}
\newcommand{\dkr}{\ensuremath{\partial_{k_r}}}

\newcommand{\Ds}{\ensuremath{d_s}}
\newcommand{\dsig}{\ensuremath{\partial_\sigma}}
\newcommand{\Dsig}{\ensuremath{d_\sigma}}

\newcommand{\Dt}{\ensuremath{D_t}}
\newcommand{\Dth}{\ensuremath{D_{\hat{t}}}}

\newcommand{\pt}[1]{\ensuremath{\frac{\partial #1}{\partial t}}}
\newcommand{\pz}[1]{\ensuremath{\frac{\partial #1}{\partial z}}}

\newcommand{\pzz}[1]{\ensuremath{\frac{\partial^2 #1}{\partial z^2}}}

\newcommand{\nn}{\nonumber}

\renewcommand{\*}{\ensuremath{\cdot}}
\renewcommand{\=}{\ensuremath{\, = \,}}

\newcommand{\eval}[1][]{\ensuremath{|_{_{#1}}}}
\def\xstrut{\rule{0pt}{2ex}}
\newcommand{\conj}{\ensuremath{\hspace{-.4em}~^{\strut *}}}
\newcommand{\res}{\ensuremath{^{\mbox{\scriptsize (r)}}}}
\newcommand{\ct}{\ensuremath{\hspace{-.4em}~^{\strut +}}}

\newcommand{\avg}[2][1]{\ensuremath{\overline{#2}^{~(T_{#1}, \v{X}_{#1})}}}
\newcommand{\tavg}[2][1]{\ensuremath{\overline{#2}^{~T_{#1}}}}

\newcommand{\half}{\ensuremath{\frac{1}{2}}}
\newcommand{\quart}{\ensuremath{\frac{1}{4}}}


\newcommand{\tH}{\ensuremath{ \tilde{H} }}
\newcommand{\Lw}{\ensuremath{ \tilde{L}_{\mbox{\scriptsize w}} }}
\newcommand{\Lmf}{\ensuremath{ \tilde{L}_{\mbox{\scriptsize mf}} }}
\newcommand{\tB}{\ensuremath{ \tilde{B} }}
\newcommand{\tP}{\ensuremath{ \tilde{P} }}
\newcommand{\tU}{\ensuremath{ \tilde{U} }}
\newcommand{\tV}{\ensuremath{ \tilde{V} }}
\newcommand{\tW}{\ensuremath{ \tilde{W} }}

\newcommand{\tT}[1][]{\ensuremath{\widetilde{T}_{\mbox{\scriptsize #1}}}}
\newcommand{\tX}[1][]{\ensuremath{\widetilde{X}_{\mbox{\scriptsize #1}}}}
\newcommand{\tl}{\ensuremath{\tilde{l}}}
\newcommand{\tf}{\ensuremath{\tilde{f}}}
\newcommand{\tN}{\ensuremath{\tilde{N}}}

\newcommand{\Ro}{\ensuremath{ \mathsf{Ro} }}


\newcommand{\ep}{\epsilon}
\newcommand{\be}{\beta}
\newcommand{\ga}{\gamma}
\newcommand{\de}{\delta}
\newcommand{\degr}{\ensuremath{^{\circ}}}
\renewcommand{\deg}{\degr}

\newcommand{\fn}{\ensuremath{f_0}}

\newcommand{\phib}{\ensuremath{\phi_\beta}}
\newcommand{\phig}{\ensuremath{\phi_\gamma}}
\newcommand{\phid}{\ensuremath{\phi_\delta}}

\newcommand{\vphib}{\ensuremath{\varphi_\beta}}
\newcommand{\vphig}{\ensuremath{\varphi_\gamma}}
\newcommand{\vphid}{\ensuremath{\varphi_\delta}}

\newcommand{\omb}{\ensuremath{\omega_\beta}}
\newcommand{\omg}{\ensuremath{\omega_\gamma}}
\newcommand{\omd}{\ensuremath{\omega_\delta}}

\newcommand{\omh}[1][]{\ensuremath{\hat{\omega}_{#1}}}
\newcommand{\omhb}{\ensuremath{\hat{\omega}_\beta}}
\newcommand{\omhg}{\ensuremath{\hat{\omega}_\gamma}}
\newcommand{\omhd}{\ensuremath{\hat{\omega}_\delta}}

\newcommand{\vk}[1][]{\ensuremath{\v{k}_{#1}}}
\newcommand{\vkh}[1][]{\ensuremath{\v{k}_{h#1}}}
\newcommand{\vkb}{\ensuremath{\v{k}_\beta}}
\newcommand{\vkbh}{\ensuremath{\v{k}_{\beta h}}}
\newcommand{\vkg}{\ensuremath{\v{k}_\gamma}}
\newcommand{\vkgh}{\ensuremath{\v{k}_{\gamma h}}}
\newcommand{\vkd}{\ensuremath{\v{k}_\delta}}
\newcommand{\vkdh}{\ensuremath{\v{k}_{\delta h}}}

\newcommand{\kb}{\ensuremath{k_\beta}}
\newcommand{\kd}{\ensuremath{k_\delta}}

\newcommand{\lb}{\ensuremath{l_\beta}}
\renewcommand{\lg}{\ensuremath{l_\gamma}}
\newcommand{\ld}{\ensuremath{l_\delta}}

\newcommand{\mb}{\ensuremath{m_\beta}}
\newcommand{\md}{\ensuremath{m_\delta}}

\renewcommand{\vv}{\v{v}}
\newcommand{\vu}{\v{u}}

\newcommand{\vun}[1][]{\ensuremath{\v{U}^{(#1)}_0}}
\newcommand{\vub}[1][]{\ensuremath{\v{U}^{(#1)}_\beta}}
\newcommand{\vug}[1][]{\ensuremath{\v{U}^{(#1)}_\gamma}}
\newcommand{\vud}[1][]{\ensuremath{\v{U}^{(#1)}_\delta}}

\newcommand{\vvn}[1][]{\ensuremath{\v{V}^{(#1)}_0}}
\newcommand{\vvb}[1][]{\ensuremath{\v{V}^{(#1)}_\beta}}
\newcommand{\vvg}[1][]{\ensuremath{\v{V}^{(#1)}_\gamma}}
\newcommand{\vvd}[1][]{\ensuremath{\v{V}^{(#1)}_\delta}}
\newcommand{\vvw}[2][1]{\ensuremath{\v{V}^{(#1)}_{#2}}}

\newcommand{\un}[1][]{\ensuremath{U^{(#1)}_0}}
\newcommand{\vn}[1][]{\ensuremath{V^{(#1)}_0}}
\newcommand{\wn}[1][]{\ensuremath{W^{(#1)}_0}}
\newcommand{\ub}[1][]{\ensuremath{U^{(#1)}_\beta}}
\newcommand{\vb}[1][]{\ensuremath{V^{(#1)}_\beta}}

\newcommand{\psig}{\ensuremath{\psi_{\mbox{\scriptsize g}}}}

\newcommand{\sub}[1]{_{\mbox{\scriptsize #1}}}

\newcommand{\w}[1][]{\ensuremath{W^{(1)}_{#1}}}
\newcommand{\Wn}[1][]{\ensuremath{W^{(#1)}_0}}
\newcommand{\Wb}[1][]{\ensuremath{W^{(#1)}_\beta}}
\newcommand{\Wg}[1][]{\ensuremath{W^{(#1)}_\gamma}}
\newcommand{\Wd}[1][]{\ensuremath{W^{(#1)}_\delta}}

\newcommand{\bn}[1][]{\ensuremath{B^{(#1)}_0}}
\newcommand{\bb}[1][]{\ensuremath{B^{(#1)}_\beta}}
\newcommand{\bg}[1][]{\ensuremath{B^{(#1)}_\gamma}}
\newcommand{\bd}[1][]{\ensuremath{B^{(#1)}_\delta}}

\newcommand{\thbar}[1][]{\ensuremath{\bar{\Theta}^{(#1)}}}
\newcommand{\thn}[1][]{\ensuremath{\Theta^{(#1)}_0}}
\newcommand{\thb}[1][]{\ensuremath{\Theta^{(#1)}_\beta}}
\newcommand{\thg}[1][]{\ensuremath{\Theta^{(#1)}_\gamma}}
\newcommand{\thd}[1][]{\ensuremath{\Theta^{(#1)}_\delta}}

\newcommand{\pn}{\ensuremath{P_0}}
\newcommand{\pb}[1][]{\ensuremath{P^{(#1)}_\beta}}

\newcommand{\pibar}[1][]{\ensuremath{\bar{\Pi}^{(#1)}}}
\newcommand{\pin}[1][]{\ensuremath{\Pi^{(#1)}_0}}
\newcommand{\pib}[1][]{\ensuremath{\Pi^{(#1)}_\beta}}

\newcommand{\Rbar}{\ensuremath{\bar{R}^{(0)}}}

\newcommand{\Tbgd}[1][]{\ensuremath{T_{\beta\gamma\delta}^{(#1)}}}
\newcommand{\Tb}[1][]{\ensuremath{T_{\beta}^{(#1)}}}
\newcommand{\Mb}{\ensuremath{M_\beta}}
\newcommand{\Zb}[1][]{\ensuremath{Z_\beta^{(#1)}}}
\newcommand{\Zbp}[1][]{\ensuremath{Z_\beta^{(#1)+}}}
\newcommand{\zb}[1][]{\ensuremath{\zeta_\beta^{(#1)}}}
\newcommand{\zg}[1][]{\ensuremath{\zeta_\gamma^{(#1)}}}
\newcommand{\zd}[1][]{\ensuremath{\zeta_\delta^{(#1)}}}
\newcommand{\zw}[2][1]{\ensuremath{\zeta_{#2}^{(#1)}}}
\newcommand{\zn}{\ensuremath{\zeta_0}}
\newcommand{\Rb}{\ensuremath{R_\beta}}

\newcommand{\xibp}[1][1]{\ensuremath{\xi_\beta^{(#1)+}}}
\newcommand{\xib}[1][1]{\ensuremath{\xi_\beta^{(#1)}}}
\newcommand{\vcgh}{\ensuremath{\hat{\v{c}}_{g}}}
\newcommand{\vcghb}{\ensuremath{\hat{\v{c}}_{g\beta}}}
\newcommand{\vcghbh}{\ensuremath{\hat{\v{c}}_{g\beta h}}}
\newcommand{\vcgb}{\ensuremath{\v{c}_{g\beta}}}
\newcommand{\vcgg}{\ensuremath{\v{c}_{g\gamma}}}
\newcommand{\vcgd}{\ensuremath{\v{c}_{g\delta}}}
\newcommand{\vcg}[1][]{\ensuremath{\v{c}_{g #1}}}
\newcommand{\cgzb}{\ensuremath{c_{g\beta z}}}
\newcommand{\cgz}[1][]{\ensuremath{c_{g z #1}}}

\newcommand{\cglambda}[1][]{\ensuremath{c_{g \lambda #1}}}
\newcommand{\cgphi}[1][]{\ensuremath{c_{g \phi #1}}}
\newcommand{\cgr}[1][]{\ensuremath{c_{g r #1}}}

\newcommand{\cgxhb}{\ensuremath{\hat{c}_{g\beta x}}}
\newcommand{\cgyhb}{\ensuremath{\hat{c}_{g\beta y}}}
\newcommand{\cgzhb}{\ensuremath{\hat{c}_{g\beta z}}}

\newcommand{\cgzg}{\ensuremath{c_{g\gamma z}}}
\newcommand{\cgzd}{\ensuremath{c_{g\delta z}}}
\newcommand{\cgzs}{\ensuremath{c_{gz}}}

\newcommand{\Eb}[1][1]{\ensuremath{E_\beta^{(#1)}}}
\newcommand{\En}{\ensuremath{E_0}}
\newcommand{\E}[1][]{\ensuremath{E_{#1}}}
\newcommand{\Ab}[1][1]{\ensuremath{\A_\beta^{(#1)}}}
\newcommand{\vpbh}[1][1]{\ensuremath{\v{p}^{(#1)}_{\be h}}}
\newcommand{\vpb}[1][1]{\ensuremath{\v{p}^{(#1)}_{\be}}}

\newcommand{\Pn}{\ensuremath{\pi_0}}
\newcommand{\PVn}{\ensuremath{\Pi_0}}

\newcommand{\Ap}[1][]{\ensuremath{A_{#1}^+}}
\newcommand{\Am}[1][]{\ensuremath{A_{#1}^-}}
\newcommand{\Apm}[1][]{\ensuremath{A_{#1}^\pm}}
\newcommand{\Cp}[1][]{\ensuremath{C_{#1}^+}}
\newcommand{\Cm}[1][]{\ensuremath{C_{#1}^-}}
\newcommand{\Cpm}[1][]{\ensuremath{C_{#1}^\pm}}
\newcommand{\Abgd}[1][]{\ensuremath{A_{\be\ga\de}^{#1}}}
\newcommand{\Bbgd}[1][]{\ensuremath{B_{\be\ga\de}^{#1}}}
\newcommand{\coeff}[2][123]{\ensuremath{A_{#1}^{#2}}}
\newcommand{\varcoeff}[2][123]{\ensuremath{B_{#1}^{#2}}}

\newcommand{\chit}{\ensuremath{\chi_T}}
\newcommand{\chiz}{\ensuremath{\chi_Z}}
\newcommand{\vchix}{\ensuremath{\v{\chi}_X}}

\newcommand{\cp}{\ensuremath{c_{\mbox{\scriptsize p}}}}
\newcommand{\cv}{\ensuremath{c_{\mbox{\scriptsize V}}}}

\newcommand{\Lh}{\ensuremath{L_{\mbox{\scriptsize h}}}}



\renewcommand{\A}{\ensuremath{\mathcal A}}
\newcommand{\B}{\ensuremath{\mathcal B}}
\renewcommand{\C}{\ensuremath{\mathcal C}}
\newcommand{\mathcalD}{\ensuremath{\mathcal D}}
\renewcommand{\F}{\ensuremath{\mathcal F}}
\newcommand{\G}{\ensuremath{\mathcal G}}
\renewcommand{\H}{\ensuremath{\mathcal H}}
\renewcommand{\J}{\ensuremath{\mathcal J}}
\renewcommand{\L}{\ensuremath{\mathcal L}}
\renewcommand{\K}{\ensuremath{\mathcal K}}
\newcommand{\M}{\ensuremath{\mathcal M}}
\renewcommand{\N}{\ensuremath{\mathcal N}}
\newcommand{\NN}{\ensuremath{\mathbb{N}}}
\newcommand{\NNn}{\ensuremath{\mathbb{N}_0}}
\renewcommand{\O}{\ensuremath{\mathcal O}}
\renewcommand{\P}{\ensuremath{\mathcal P}}
\newcommand{\Q}{\ensuremath{\mathcal Q}}
\newcommand{\R}{\ensuremath{\mathsf R}} 
\newcommand{\RR}{\ensuremath{\mathbb R}}
\renewcommand{\S}{\ensuremath{\mathcal S}}
\newcommand{\T}{\ensuremath{\mathcal T}}
\newcommand{\U}{\ensuremath{\mathcal U}}
\renewcommand{\V}{\ensuremath{\mathcal V}}
\newcommand{\WW}{\ensuremath{\mathcal W}}
\newcommand{\X}{\ensuremath{\mathcal X}}
\newcommand{\Y}{\ensuremath{\mathcal Y}}
\newcommand{\Z}{\ensuremath{\mathcal Z}}

\newcommand{\qmq}[1]{\ensuremath{\quad\quad\mbox{#1}\quad\quad}}
\newcommand{\qand}{\ensuremath{\quad\quad\mbox{and}\quad\quad}}
\newcommand{\qwith}{\ensuremath{\quad\quad\mbox{with}\quad\quad}}


\newcommand{\req}[1]{(\ref{#1})}
\renewcommand{\eqref}[1]{Eq. (\ref{#1})}
\newcommand{\eqrefs}[2]{Eqs. (\ref{#1}) and (\ref{#2})}
\newcommand{\eqrefr}[2]{Eqs. (\ref{#1}) to (\ref{#2})}
\newcommand{\peqref}[1]{(Eq. \ref{#1})}
\newcommand{\peqrefs}[2]{(Eqs. \ref{#1} and \ref{#2})}
\newcommand{\peqrefr}[2]{(Eqs. \ref{#1} to \ref{#2})}


\newcommand{\lhs}{left-hand side }
\newcommand{\rhs}{right-hand side }
\newcommand{\ie}{i.e.~}
\newcommand{\eg}{e.g.~}
\newcommand{\cf}{cf.~}

\newcommand{\freq}{frequency }
\newcommand{\freqn}{frequency}
\newcommand{\freqs}{frequencies }
\newcommand{\frqesn}{frequencies}
\newcommand{\BVF}{Brunt-V\"ais\"al\"a frequency }


\newcommand{\eps}{\varepsilon}

\def\ex{\v{e}_{x}}
\def\ey{\v{e}_{y}}
\def\ez{\v{e}_{z}}
\def\er{\v{e}_r}
\def\el{\v{e}_\lambda}
\def\ephi{\v{e}_\phi}

\def\kl{k_\lambda}
\def\kp{k_\phi}
\def\kr{k_r}

\def\vr{\v{r}}
\def\vkv{\v{k}}

\def\pvh{\v{p}_h}
\def\uv{\v{u}}
\def\vv{\v{v}}
\def\wv{\v{w}}
\def\xv{\v{x}}
\def\Xv{\v{X}}
\def\kv{\v{k}}
\def\kvb{\v{k}_\beta}
\def\kvhb{\v{k}_{h,\beta}}
\def\khb{k_{h,\beta}}
\def\kvp{\dot{\v{k}}}
\def\cv{\v{c}}
\def\cg{\v{c}_g}
\def\cgb{\v{c}_{g,\beta}}
\def\cgh{\v{\hat{c}}_g}
\def\cgmh{\v{\hat{c}}_\gamma}
\def\cgmb{\v{\hat{c}}_{\beta,\gamma}}
\def\cph{\v{c}_p}
\def\tv{\v{\mathcal{\tau}}}
\def\Vv{\v{V}}
\def\Uv{\v{U}}
\def\Fv{\v{F}}
\def\Tv{\v{T}}
\def\Mv{\v{M}}

\def\vva{\langle \vv \rangle}
\def\vua{\langle \uv \rangle}
\def\ua{\langle u \rangle}
\def\vma{\langle v \rangle}
\def\wa{\langle w \rangle}
\def\ba{\langle b \rangle}
\def\dta{\langle \delta\theta \rangle}
\def\pia{\langle \pi \rangle}

\newcommand{\rhobar}{\overline{\rho}}

\def\tbar{\overline{\theta}}

\def\cgh{\v{\hat{c}}_g}

\def\Ac{\mathcal{A}}
\def\Nc{\mathcal{N}}

\def\Gc{\mathcal{G}}
\def\Hc{\mathcal{H}}

\newcommand{\vFc}{{\boldsymbol{\mathcal{F}}}}
\newcommand{\vGc}{{\boldsymbol{\mathcal{G}}}}
\newcommand{\vHc}{{\boldsymbol{\mathcal{H}}}}

\def\oh{\hat{\omega}}

\def\Rhobarnull{{\overline{R}^{(0)}}}

\newcommand{\Pbarnull}{\overline{P}^{(0)}}

\newcommand{\pibarnull}{{\overline{\Pi}^{(0)}}}

\def\tbarnull{{\overline{\Theta}^{(0)}}}
\def\tbaralpha{\overline{\Theta}^{(\alpha)}}

\def\Vv{\v{V}}
\def\Uv{\v{U}}

\newcommand{\Ord}{\mathcal{O}}

\newcounter{mycount}
\newcounter{mycount2}

\newcommand{\myDiv}[1]{\nabla\cdot#1}

\newcommand{\mySum}[3]{
  \sum\limits_{#1=#2}^{#3}
}

\newcommand{\fld}[3]{{#1}_{#2}^{(#3)}}
\newcommand{\Vfld}[2]{\Vv_{#1}^{(#2)}}
\newcommand{\Ufld}[2]{\Uv_{#1}^{(#2)}}
\newcommand{\Wfld}[2]{W_{#1}^{(#2)}}

\section{Introduction}
\label{sec:intro}

Internal gravity waves (IGWs) play a significant role in distributing energy and momentum throughout the atmosphere. Being generated through the perturbation of balanced flow states in stratified environments (e.g. convection, flow over topography, flow instabilities, wave turbulence, etc.) they may propagate over large distances and deposit their energy, generate turbulence and modify the mean-flow dynamics in general far away from their source \citep{Fritts2003, Alexander2010, Sutherland2010, Nappo2013, Williams2017, Sutherland2019, Achatz2022}. While propagating, they may exchange energy among themselves, transiently interact with mean flows, or influence the transport of chemical species. Even though their generation regions are mostly located in the troposphere, their effects are strongest in the middle atmosphere \citep{Lindzen1981, Kim2003a, Sigmond2010}. However, these major effects may couple to and thus impact for instance tropospheric weather patterns or climate conditions \citep{Scaife2005, Scaife2012, Sigmond2010}.

In state-of-the-art general circulation models (GCMs) or numerical weather prediction (NWP) models, gravity waves are only partially resolved and thus need parameterization \citep[e.g.][]{Kim2003a, Holt2016}. Even when entering the global kilometer-resolving regime, their effects are not entirely represented \citep{Kruse2022, Polichtchouk2022, Polichtchouk2023}. IGW parameterizations commonly rely on WKBJ theories \citep{Bretherton1966, Grimshaw1975, Achatz2017}, incorporating some major simplifications \citep{Lindzen1981, Medvedev1995, Warner1996, Hines1997, Hines1997a, Lott1997, Alexander1999, Scinocca2003, Orr2010, Lott2013}. Most notably, there are three assumptions that are usually made and shall be considered in this work. Firstly, local horizontal homogeneity is assumed in the so-called single-column approximation, i.e. horizontal gradients of the resolved flow are neglected, which leads horizontal wave numbers to be invariant during propagation. Furthermore, responses of the resolved flow due to the horizontal finiteness of an IGW field are neglected. Secondly, a steady-state assumption is adopted such that IGWs instantly propagate through the atmosphere, leading to a neglect of all transient propagation effects. Thirdly, it is commonly assumed that the resolved flow is (approximately) in hydrostatic and geostrophic balance, implying a reduced formula for the resolved-flow response to IGWs. In the extratropics, this can lead to significant modifications in the IGW forcing when the flow is imbalanced.

Recent investigations \added[id=GSV]{suggested} that the mentioned effects can have important impacts on modeled flows \citep{Sato2009, Boloni2016, Ehard2017, Wei2019}. \added[id=GSV]{In particular, \cite{Boloni2016} showed, that including the time dependence of the gravity wave propagation significantly improves the estimated wave drag in idealized simulations. \cite{Sato2009} found that three-dimensional propagation pathways can impact the structure of the stratosphere and mesosphere using wave resolving general circulation models. Similarly, \cite{Ehard2017} highlight the importance of horizontal gravity wave propagation for the dynamics of the Antarctic polar night jet by analyzing various observations and reanalyses. Finally, \cite{Wei2019} showed that describing the wave momentum fluxes with a direct formulation can improve the simulated wave impact on the resolved flow.} Several studies have successfully relaxed some of the simplifications \added[id=GSV]{in model parameterizations} \citep[e.g.][]{Muraschko2015, Wilhelm2018, Quinn2020}. In particular, \cite{Boloni2021} and \cite{Kim2021} presented a novel Lagrangian IGW parameterization, the Multi-Scale Gravity-Wave Model (MS-GWaM). It is built on a weakly non-linear WKBJ theory, including transient wave-mean flow interactions. The model has been implemented in a single-column mode into a state-of-the-art weather forecast and climate code. It will therefore be referred to as MS-GWaM-1D throughout this manuscript. Here, we present MS-GWaM-3D, an extension of MS-GWaM-1D, which models the full 3D transient propagation of IGWs and their forcing of the general (cf. balanced) resolved flow. \added[id=GSV]{While both the columnar approximation and assumptions on the resolved flow are relaxed in MS-GWaM-3D, we shall focus on the effects of the horizontal propagation in this study. In particular, the modified refraction behavior and resolved flow response in the statistical mean are investigated. Indeed we find that many aspects show an increased reaslism of the wave dynamics. However, our results also underline the challenges left for future studies.} Ultimately, we aim to improve the representation of the parameterized IGW processes while keeping simulations numerically efficient.

This paper is structured as follows. First, a brief recapitulation of the underlying WKBJ theory for transient, 3-dimensional IGWs (Sec. \ref{sec:theory}) is presented. It is then followed by the description of the employed ray tracing techniques (Sec. \ref{sec:raytracing}). Lastly, results from  simulations with MS-GWaM-3D (Sec. \ref{sec:results}) are visualized and discussed. The manuscript then closes with some concluding remarks on the achievements and challenges of ray tracing parameterization (Sec. \ref{sec:conclusions}).

\section{Nonlinear, 3-dimensional and transient internal gravity waves}
\label{sec:theory}

Albeit generally 3-dimensional and transient, internal gravity waves are commonly parameterized using both the single-column and the steady-state approximations. Here, we attempt to relax both these assumptions and build a 3-dimensional and transient parameterization. \added[id=GSV]{An underlying multiscale theory has been presented by \cite{Achatz2010, Achatz2017, Achatz2023b} and thus we restrict ourselves to summarizing the results here for brevity.}

\subsection{Wave evolution}
\label{subsec:BG-modulation-equations}

We follow the analysis of \cite{Bretherton1966, Grimshaw1974, Hasha2008, Achatz2017, Achatz2022, Achatz2023b} and summarize the Bretherton-Grimshaw modulation equations for a quasi-monochromatic wave train on a sphere. \added[id=GSV]{In particular the wave and mean-flow dynamics are derived by means of a WKBJ expansion of the compressible Euler equations. We extend the existing theories by} the inclusion of the atmospheric density scale height to improve the realism in the non-hydrostatic regime. \added[id=GSV]{Most notably, the scale height correction plays an important role in the course of vertical wave reflection, where vertical wave numbers become small with respect to the inverse scale height, $\Gamma$. We then arrive at the} well-known dispersion relation,
\begin{align}
    \label{eq:dispersion}
    \omega(\v{r}, t) &= \Omega(\v{r}, t, \vk) \nn \\
    &= \vk \* \v{U} \pm \sqrt{\frac{f^2 (k_r^2 + \Gamma^2) + N^2(k_\lambda^2 + k_\phi^2)}{K^2}},
\end{align}
with the position vector, $\v{r}= \lambda\v{e}_\lambda + \phi\v{e}_\phi + r \er$, and the wave vector, $\vk =\kl\el + \kp\ephi + \kr\er$. Here $\lambda,\phi$ and $r$ are the standard geographical spherical coordinates, and $\el,\ephi$ and $\er$ are the corresponding unit vectors. Note that the mean-flow velocity $\v{U}(\v{r}, t) = U\el + V\ephi$, the Coriolis frequency, $f=f(\v{r})$, and the buoyancy frequency, $N=N(\v{r}, t)$, are functions of space and time as indicated. For convenience, we have defined the total wave vector squared as $K^2 = \vk\*\vk + \Gamma^2$, with the scale height correction $\Gamma = \Gamma(\v{r}, t)$. Finally, we would like the reader to note that the theory is written in spherical coordinates. The corresponding eikonal equations and the group velocities, $\v{c}_g$, for an individual wave component then follow the relations
\begin{align}
    \label{eq:eikonal}
    (\dt + \v{c}_g\*\nabla_{\vr})\vk &= \dot{\vk} = -\nabla_{\vr}\Omega, &
    \v{c}_g &= \nabla_{\vk}\Omega.
\end{align}
Here we denote the gradients with respect to position and wave number by $\nabla_{\vr}$ and $\nabla_{\vk}$, respectively. Having the application to a ray tracing scheme in mind, we indicate by a dot the derivative along characteristics, so-called rays, with their positions denoted by $\v{r}(t)$ so that
\begin{align}
    \label{eq:group-velocity}
    \dot{\v{r}} = \v{c}_g.
\end{align}
Rays follow the local group velocity, and hence the propagation of wave energy. In particular, the prognostic equations for the zonal, meridional, and radial wave numbers, $(k_\lambda, k_\phi, k_r)$, in spherical coordinates read
\begin{align}
    \label{eq:eikonal-sphere:1}
    \dot{k_\lambda} =
      &-\frac{1}{r\cos\phi} \left(\v{k}\*\dlambda\v{U}  + \frac{|\v{k}_h|^2}{2\omh K^2}\left[\dlambda N^2 - \frac{N^2 - f^2}{K^2}\dlambda\Gamma^2\right]\right)
      - \frac{k_\lambda}{r}c_{gr} + \frac{k_\lambda\tan\phi}{r}c_{g\phi},\\
    \label{eq:eikonal-sphere:2}
    \dot{k_\phi} =
      &-\frac{1}{r} \left(\v{k}\*\dphi\v{U} + \frac{|\v{k}_h|^2}{2\omh K^2}\left[\dphi N^2 + \frac{k_r^2 + \Gamma^2}{|\v{k}_h|^2}\dphi f^2 - \frac{N^2 - f^2}{K^2}\dphi \Gamma^2\right]\right)
      - \frac{k_\phi}{r}c_{gr} - \frac{k_\lambda\tan\phi}{r}c_{g\lambda},\\
    \label{eq:eikonal-sphere:3}
    \dot{k_r} =
      &- \left(\v{k}\*\dr\v{U} + \frac{|\v{k}_h|^2}{2\omh K^2}\left[\dr N^2 - \frac{N^2 - f^2}{K^2}\dr \Gamma^2\right]\right)
      + \frac{k_\lambda}{r}c_{g\lambda} + \frac{k_\phi}{r}c_{g\phi}.
\end{align}
Note that the metric terms ensure the propagation on great circles in the case of a uniform atmosphere at rest, as laid out in detail by \cite{Hasha2008}\footnote{Revisiting past studies we recognize that \cite{Ribstein2015} and \cite{Ribstein2016} omit some of the important metric terms so that their results should be interpreted with care.}. Note that the representation in spherical coordinates comes with the drawback of a pole problem in the above relations when propagating very close to the pole. Practically, singularities due to division by zero, however, do not pose a serious problem as they only occur in close vicinity of the poles. To avoid potential problems, we do not consider wave generation where $|\phi|>85^\circ$ (see Sec. \ref{sec:raytracing}\ref{subsec:sources-sinks}).

Finally, for the locally monochromatic field, its amplitude is governed by the wave-action equation
\begin{align}
    \label{eq:wa-3d-conservation}
    0 &= \dt \A + \nabla_{\v{r}}\*(\v{c}_g \A) + S,
\end{align}
for the wave-action density $\A$ in physical space, where $S$ denotes \added[id=GSV]{sources and sinks of wave action, such as the} wave saturation as described below. Wave-wave interactions between IGWs as well as interactions between IGWs and the (subgrid-scale) geostrophic modes (GM) are neglected, and thus the wave action conservation is valid for any overlapping wave fields, provided their amplitudes are sufficiently weak. Numerous processes, e.g. wave refraction, can lead to rays crossing in physical space at so-called caustics, so that the assumption of local monochromaticity can break down in the course of the integration of the eikonal equations above. To avoid numerical instabilities due to this issue, we introduce the spectral wave-action density $\N=\N(\v{r}, \v{k},t) = \sum_j \A_j(\v{r},t) \delta\left[\vk - \vk[j](\v{r},t)\right]$ in phase space, \citep[c.f.][and references therein]{Muraschko2015, Achatz2022} where the index $j$ indicates any member of a possibly infinitely large and infinitely dense set of locally monochromatic fields that are being superposed. The resulting phase-space wave-action conservation then reads
\begin{align}
    \label{eq:wa-6d-conservation}
    0 &= \dt\N + \nabla_{\v{r}}\*(\v{c}_g\N) + \nabla_{\vk}\*(\dot{\v{k}}\N) + \S\\
      &= (\dt + \v{c}_g\*\nabla_{\v{r}} + \dot{\v{k}}\*\nabla_{\vk})\N + \S\nn.
\end{align}
Note that due to the non-divergence of the six-dimensional phase-space velocity,
\begin{equation}
    \label{eq:vpvel_nondiv}
    0=\nabla_{\v{r}}\*\vcg + \nabla_{\v{k}}\*\dot{\v{k}},
\end{equation}
the flux and the transport formulations in \eqref{eq:wa-6d-conservation} are equivalent. \added[id=GSV]{Again, $\S$ denotes both sources and the} dissipation of phase-space wave action. Thus, the phase-space wave-action density, $\N$, is conserved along wave propagation paths up to the wave dissipation \added[id=GSV]{and generation}. Moreover, the non-divergence \peqref{eq:vpvel_nondiv} of the phase-space velocity also implies that any phase-space volume following its rays conserves its volume content. \added[id=GSV]{As a consequence, rays cannot cross in phase space and caustics do not occur.} Summarizing, \eqrefr{eq:group-velocity}{eq:eikonal-sphere:3} and \eqref{eq:wa-6d-conservation} form the basis for the forward ray-tracing of gravity waves.

\subsection{Wave impact on the mean flow}
\label{subsec:wave_fluxes}

To account for the impact of the wave perturbations on the mean flow, we note that the IGW-perturbation fluxes enter the mean flow tendencies as
\begin{align}
    \label{eq:mean-flow-impact}
    \left.\dt \v{U}\right|_{IGW} =
        &-\frac{1}{\bar{\rho}}\nabla\*(\bar{\rho}\left\langle\v{v}'\v{u}'\right\rangle)\\
        &+ \frac{f}{g}\v{e}_r\times\left\langle \v{u}'b'\right\rangle,\nn\\
    \label{eq:mean-flow-impact-t}
    \left.\dt \theta\right|_{IGW} = 
        &-\nabla\*\left\langle \v{u}'\theta'\right\rangle,
\end{align}
where $\v{e}_r$ is the radial unit vector, $g$ is Earth's gravity, and $\bar{\rho}$ and $\bar{\theta}$ are the reference density and potential temperature \citep{Achatz2017, Achatz2023b}. Moreover, $\v{v}'$, $b'$, and $\theta'$ denote the IGW velocity, buoyancy, and potential-temperature perturbations, and $\v{u}'$ the horizontal component of $\v{v}'$. The brackets, $\left\langle.\right\rangle$, represent the phase average of the wave perturbations. Using the dispersion and polarization relations, they may be expressed as
\begin{align}
    \label{eq:fluxes:momentum}
    \bar{\rho}\left\langle \v{v}'\v{u}'\right\rangle &= \int \left[\frac{
        \omh^2\cgh\vkh\N + f^2(\v{e}_r\times\cgh)(\v{e}_r\times\vkh\N)
        }{\omh^2 - f^2}\right]d^3k,\\
    \label{eq:fluxes:temperature}
    \left\langle \v{u}'\theta'\right\rangle
    &= \int \left[
        \frac{\bar{\theta}}{g\bar{\rho}} \hat{c}_{gr} \frac{fN^2}{\omh^2 - f^2} \v{e}_r\times\vkh\N
        \right]d^3k,\\
    \label{eq:fluxes:buoyancy}
    \frac{f}{g}\er\times\left\langle \v{u}'b'\right\rangle 
    &= \frac{f}{\bar{\theta}}\er\times\left\langle \v{u}'\theta'\right\rangle,
\end{align}
where we denote the horizontal wave vector as $\vkh = k_\lambda\el + k_\phi\ephi$, and where $\cgh = \vcg - \v{U}$ is the intrinsic group velocity. Note that the elastic term, $\frac{f}{g}\v{e}_r\times\left\langle \v{u}'b'\right\rangle$, and the potential-temperature flux divergence, $\nabla\*\left\langle \v{u}'\theta'\right\rangle$, \added[id=GSV]{are negligible where the Coriolis frequency, $f$, is small with respect to the intrinsic wave frequency, $|f|\ll|\omh|$. Thus, the two terms} are weak for non-hydrostatic waves\added[id=GSV]{ in the mid to high frequency range}. \added[id=GSV]{In addition, the momentum flux divergence may then be approximated with the pseudo}momentum flux divergence, $\nabla\*(\bar{\rho}\left\langle\v{v}'\v{u}'\right\rangle) \approx \nabla\*\int \cgh\vkh\N d^3k$. \added[id=GSV]{Interestingly, the corresponding} flux, often termed Eliassen-Palm flux, \added[id=GSV]{also determines the evolution of the mean-flow potential vorticity and the geostrophically and hydrostatically balanced mean flow even without any spectral restriction. It is thus often used in IGW parameterizations \citep[e.g.][]{Warner2001,Orr2010,Boloni2021}. For a brief comparison of the two approaches, see Appendix A. Recently, \cite{Wei2019} showed that the direct flux formulation more accurately represents the impact of an internal wave packet on the mean flow. Hence, we use the full set of fluxes shown in \eqrefr{eq:mean-flow-impact}{eq:fluxes:buoyancy}.}\\
Combined with the eikonal equations \peqrefr{eq:eikonal-sphere:1}{eq:eikonal-sphere:3}, wave propagation \peqref{eq:group-velocity}, and wave-action conservation \peqref{eq:wa-6d-conservation}, these relations form a closed prognostic system for transient wave propagation. Note that the equations are energy-conserving wherever both wave sources and dissipation are zero (not shown explicitly).

\section{Gravity wave ray tracing}
\label{sec:raytracing}

As pointed out in section \ref{sec:theory}, the subgrid-scale gravity wave dynamics are natively described on the rays of the corresponding gravity wave propagation. To include both the transience and the horizontal wave propagation, one may employ various techniques to obtain a numerical solution. Ray tracing has gained significant attention recently as an efficient and accurate method for simulating wave fields \citep{Marks1995, Muraschko2015, Amemiya2016, Voelker2020, Boloni2021}. Here, we build on the implementations of \cite{Boloni2021} and \cite{Kim2021} and extend the approach by the horizontal wave propagation and the full flux calculation \peqrefr{eq:mean-flow-impact}{eq:mean-flow-impact-t}. While we present a broad overview in this section, we would like to refer the reader to Appendix A for additional details. 

\subsection{Representation of rays as phase-space ray volumes}
\label{subsec:repray}

One of the major challenges for ray tracing is the problem of caustics and the subsequent ambiguity of the notion of wave amplitudes and wave energy of a local spectrum. As laid out above, it is possible to reformulate the wave action conservation as a transport equation in the 6-dimensional phase space. The non-divergence of the phase-space velocity \peqref{eq:vpvel_nondiv} implies that any phase-space volume following its rays conserves its volume content, i.e.
\begin{align}
    0 &= d_t\left(dx dy dz \,dk_x dk_y dk_z\right)\nn\\
      &= d_t\left(r^2 \cos\phi d\lambda d\phi dr \,dk_\lambda dk_\phi dk_r\right).
\end{align}
where $(x,y,z)$ are the Cartesian coordinates, and $(k_x,k_y,k_z)$ are the wave-vector components in these coordinates.
In Cartesian coordinates one has even
\begin{align}
    0 &= \dx c_{gx} + \partial_{k_x} \dot{k}_x\\
    0 &= \dy c_{gy} + \partial_{k_y} \dot{k}_y\\
    0 &= \dz c_{gz} + \partial_{k_z} \dot{k}_z
\end{align}
so that
\begin{equation}
    \label{eq:vpvol_cons_cart}
    0 
    = d_t \left(dx dk_x\right) 
    = d_t \left(dy dk_y\right) 
    = d_t \left(dz dk_z\right).
\end{equation}
Such a simple splitting \added[id=GSV]{relies on the independence of the spatial coordinates and} does not exist in spherical coordinates. The approach taken in MS-GWaM, for the time being, is to divide the phase-space content with non-zero wave-action density, $\N$, into finite-size small ray volumes, with extent $\Delta (\lambda,\phi,r,k_\lambda,k_\phi,k_r)$. Within the spanned ray volume, the wave action density, $N$, is assumed constant. The ray volume content is then propagated according to \eqrefr{eq:group-velocity}{eq:eikonal-sphere:3} following a central carrier ray. The change in physical extent, $\Delta(\lambda,\phi,r)$, is determined using rays at the faces of the ray volumes carrying identical wave properties $(k_\lambda,k_\phi,k_r)$ as in the central carrier ray but being exposed to deviations in the background fields. In particular, the ray-volume extent is approximated as locally Cartesian by
\begin{align}
    \Delta (x,y,z) &= (r \cos\phi \Delta\lambda, r \Delta\phi, \Delta r),\\
    \Delta (k_x, k_y, k_z) &= \Delta (k_\lambda, k_\phi, k_r) \label{eq:dkxdkl}
\end{align}
where the new $\Delta (k_x,k_y,k_z)$ is determined by \eqref{eq:vpvol_cons_cart}, and then inverted using \eqref{eq:dkxdkl} to finally obtain the new $\Delta(k_\lambda,k_\phi,k_r)$. Note that this procedure, albeit converging for infinitesimally small ray volumes, implies the following simplification. In a uniform atmosphere at rest, where the group velocity differences between the opposing ray-volume faces are exactly zero, the ray-volume extent in spherical coordinates, $\Delta(\lambda, \phi, r)$, is exactly conserved. This implies, however, that any ray volume propagating polewards must shrink in the tangent-linear extent $\Delta x$ and expand in spectral width $\Delta k_\lambda$ (and vice versa for southward propagation). Although this effect may be small in regions sufficiently separated from the pole, we plan to apply an advanced method to remove this \added[id=GSV]{potential} problem to the next version of MS-GWaM.

\subsection{Coupling of Lagrangian particles to the Eulerian mean flow}
\label{subsec:coupling}

The coupling between the large-scale flow and the wave perturbations necessitates linking the Lagrangian wave representation to the Eulerian model grid. For the evaluation of the modulation equations, we linearly interpolate all necessary mean fields and their gradients to the ray-volume center or faces using a first-order Taylor expansion. In particular, the group velocities are calculated at the physical boundaries of the ray volume (see App. \ref{app:interpolation}). This serves a dual purpose: the mean of the group velocities on opposing ray-volume faces may be used to advance the location of the central carrier ray, and the difference between them serves as an estimate for the compression or inflation of the ray volume in physical space. The phase-space extent is then determined as described above.\\

To account for the impact of waves on the mean flow, it is necessary to differentiate between the vertical and horizontal components, considering possibly irregular horizontal grid structures. The vertical gradient of the wave flux, i.e., $\partial_r (\bar{\rho}\left\langle w'\v{u}'\right\rangle)$ in \eqref{eq:mean-flow-impact}, 
is computed following \cite{Muraschko2015}, \cite{Boloni2016, Boloni2021}. In particular, the vertical flux associated with a grid cell boundary is integrated over all ray volumes contained in the vertically staggered cell that covers the desired cell boundary \citep[for more details, see Fig.~1a in ][]{Boloni2021}. The gradients are then computed from the centered difference of the vertical cell-boundary values. For the horizontal fluxes, the grid-cell averaged values (rather than those at the boundaries) are calculated to avoid problems with complex grid geometries. We integrate all ray volumes whose carrier rays are contained in the chosen grid cell's volume. To mend errors due to ray volumes overlapping multiple adjacent cells, the obtained flux fields are horizontally smoothed (App. \ref{app:smoothing}). The resulting fields and their divergences \peqrefs{eq:mean-flow-impact}{eq:mean-flow-impact-t} then contribute to the tendencies provided by the parametrization.

\subsection{Sources and sinks of internal gravity waves}
\label{subsec:sources-sinks}

\begin{figure*}
    \includegraphics[width=1\textwidth]{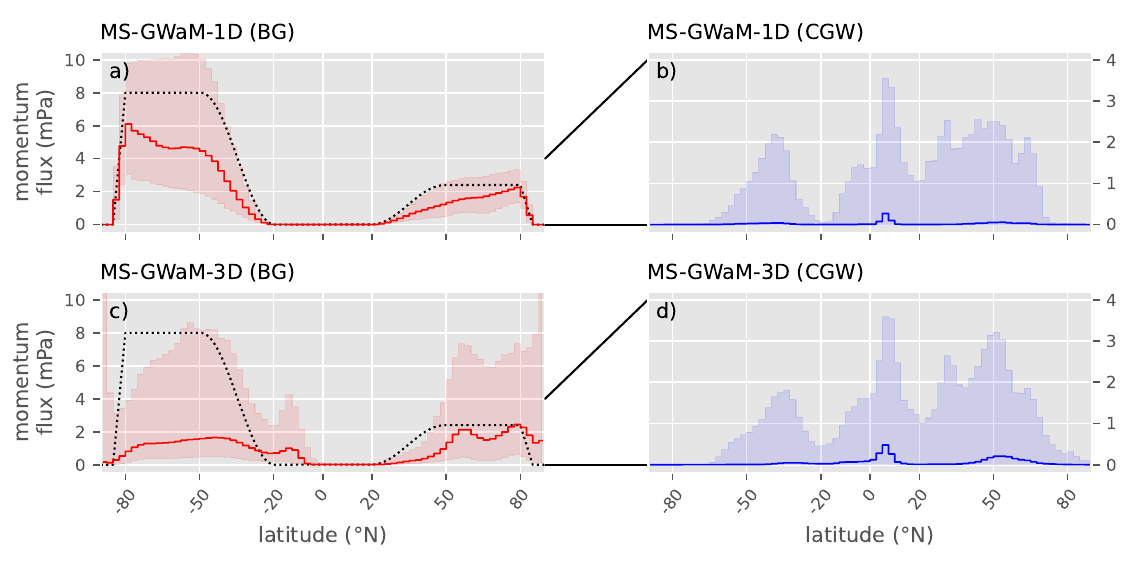}
    \caption{Zonally averaged absolute vertical fluxes of horizontal momentum at \qty{16}{\km} height for MS-GWaM-1D (a-b) and MS-GWaM-3D (c-d) for June 1991. The solid lines represent the 50th percentile (median), the surfaces show the ranges from the 10th to the 90th percentiles. Results are separated into background waves (a, c) and convectively generated waves (b, d). The black dotted lines represent the launched absolute momentum fluxes at \qty{300}{\hecto\Pa} during the summer solstice.}
    \label{fig:lower_boundary}
\end{figure*}

Similar to \cite{Boloni2021} and \cite{Kim2021}, the present implementation of MS-GWaM utilizes two sources of gravity waves as a lower boundary condition (Fig. \ref{fig:lower_boundary}). In particular, a background source in the middle to high latitudes, accounting, e.g. for IGWs emitted from jets and fronts, as well as a convective source, are employed. The former prescribes at a given launch level a seasonally varying pseudo-momentum flux with a spectral shape following the Desaubie spectrum \citep[c.f.][]{Orr2010}. At the solstice, it has a latitudinal profile with winter hemispheric fluxes of $F_w = \qty{8}{\milli\Pa}$, no fluxes for latitudes $\phi\in$[$20^\circ$S, $20^\circ$N], and summer hemispheric fluxes of $F_s = \qty{2.4}{\milli\Pa}$. No background waves are launched at latitudes higher than $|\phi|>85^\circ$. The different regimes are connected through smooth transitions. The profile oscillates in time with a sinusoidal yearly cycle. Summarizing, we employ a latitudinal profile, $\alpha(\varphi)$, and a total absolute momentum flux at the lower boundary during the Northern summer solstice, $F_{sol}$,
\begin{align}
    \alpha(\varphi) &= 
    \begin{cases}
        \min\left(1, \max\left(0, 1 - \frac{\varphi - 80}{5}\right)\right) & \mbox{if } \varphi \geq 80\\
        1, & \mbox{if } \varphi \in [50, 80)\\
        \half\left(1  - \cos\left(\pi \frac{\varphi - 20}{30}\right)\right) & \mbox{if } \varphi\in [20, 50)\\
        0 & \mbox{else}
    \end{cases},\\
    F_{sol}(\varphi) &= \alpha(\varphi)F_s + \alpha(-\varphi)F_w,
\end{align}
where $\varphi$ is the latitude in units $^\circ$N. The absolute momentum flux at the lower boundary condition then reads
\begin{align}
    F_{am} &= \half\left[\vphantom{\half}F_{sol}(\varphi)(1 + \cos\tau) \right.\nn\\
        &\qquad + \left. F_{sol}(-\varphi)(1 - \cos\tau)\vphantom{\half}\right],
\end{align}
with the non-dimensional time difference,  $\tau = 2\pi(t - t_{sol}) / \qty{1}{\y}$, relative to the time of the Northern summer solstice, $t_{sol}$. These background waves are launched \added[id=GSV]{at every time step} at a height corresponding to a pressure of \qty{300}{\hecto\Pa} (black, dotted line in Fig. \ref{fig:lower_boundary}a and c) with a momentum flux which is equally divided into 4 horizontal directions as done by \cite{Boloni2021}. \added[id=GSV]{In particular, we construct a ghost layer at the launch altitude within which ray volumes are allowed to propagate vertically only. At every launch interval, that is every \qty{60}{\s} (equivalent to the model time step), the fraction of the ray volume which has propagated out of the ghost layer, is separated and activated. Only then it will propagate freely according to the modulation equations. Ray volumes which are reflected within the ghost layer are discarded. More technical information on the launch process can be found in the work of \cite{Boloni2021}. This background} wave energy source is directionally homogeneous, with a purely latitudinal profile on the \qty{300}{\hecto\Pa} pressure surface. Additionally, convectively generated waves are considered \added[id=GSV]{with a launch spectrum} as described by \cite{Kim2021}. \added[id=GSV]{In their formulation, the areal fraction of the convective latent heat release, $\epsilon_q$, inversely controls the total entropy forcing and thus the launch momentum flux \citep{Kim2021}. Here, we increase the area fraction to $\varepsilon_q=\qty{7}{\%}$, resulting in a reduced gravity wave flux. For the convective source we couple MS-GWaM to the convection parameterization and utilize the convection strength and depth calculated within. The ray volumes are then launched at the cloud top and under consideration of the background wind.} As an example of the resulting lower boundary conditions, we show the absolute vertical momentum fluxes, defined by
\begin{align}
    \label{eq:absolute-mom-flux}
    F_{am}=\bar{\rho}\left(\left\langle w'u'\right\rangle^2 + \left\langle w'v'\right\rangle^2\right)^{\half},
\end{align}
through the surface at a constant height of \qty{16}{\km} for June 1991 (Fig. \ref{fig:lower_boundary}). The thick line presents the median of the zonally averaged fluxes during the month, and the shading around the median shows the range between the 10th and 90th percentiles, highlighting the variability in the fluxes. Even though this height is rather close to the launch altitudes, we observe that the horizontal wave propagation changes the statistics of the vertical fluxes of background waves (Fig. \ref{fig:lower_boundary}a and c). \added[id=GSV]{In particular, we note that the gravity wave transience leads to a reduction of the median absolute momentmum fluxes and a strong variability with fluxes regularly exceeding the launch flux. This can be explained by a spatially and temporally varying mean flow through which the waves propagate from the launch level to the depicted \qty{16}{km} altitude. For instance, the waves may encounter reductions and increases of their vertical group velocity, critical layers, or vertical reflection. In addition, the horizontal propagation leads to both an increased dissipation below \qty{16}{\km} as well as a horizontal redistribution, effectively changing the wave flux statistics. Comparing Fig. \ref{fig:lower_boundary}a and b we thus find a reduced median flux for MS-GWaM-3D and background waves far outside the mid to high latitude generation regions.} In contrast \added[id=GSV]{to the background source}, the convectively generated waves show rather similar statistics at this altitude between MS-GWaM-3D and 1D (Fig. \ref{fig:lower_boundary}b and d). \added[id=GSV]{Naturally, their statistics differs from the background waves as convective events are spatially and temporally rather sparse but can lead to high momentum fluxes. Thus, the spatio-temporal median is very low, while the peak momentum fluxes are comparable to the background waves.}

\added[id=GSV]{Finally, we would like to add that the launch fluxes, $F_w$ and $F_s$, as well as the area fraction, $\epsilon_q$, were tuned to achieve a realistic zonal mean mesopause. In particular, both the wind reversal and the summer hermispheric cold pole are expected to occur at approximately \qty{80}{\km} (see discussion of mean model states in Sec. \ref{sec:results}\ref{subsec:zonal-mean}). To avoid high computational costs due to optimization algorithms, short runs were considered, integrating the model for 2 weeks in both seasons. The model results, calculated for a physical range of the three tunable parameters, were then compared and the best fit was chosen. The resulting directional momentum fluxes compare in order of magnitude with satellite observations at altitudes \qtyrange[range-units=single]{30}{40}{\km} but differ in horizontal distribution \citep[][direct comparisons not shown]{Jiang2006, Ern2018, Hindley2020, Polichtchouk2022}. The latter is, however, expected given MS-GWaM incorporates a flow-independent background source.}

The two sources of wave energy are accompanied by a saturation parametrization following \cite{Boloni2016, Boloni2021} which is based on the idea of \cite{Lindzen1981} but in a spectrally integrated manner. In particular, the threshold for static instability is formulated for the wave action as follows,
\begin{align}
    \label{eq:instability-threshold}
    \int\limits_{\mathbb{R}^3} \frac{k_r^2|\vkh|^2}{\omh K^2}\N d^3k = \P < \frac{\bar{\rho}}{2}.
\end{align}
Note, that for a quasi-monochromatic wave, this relation reduces to the well-known threshold, $|B_w|^2 < N^4/k_r^2$, with $B_w$ being the amplitude of IGW buoyancy. The concept of the integrated density perturbation, $\P$, is then used in a discrete sense and rewritten as the sum over all physically overlapping ray volumes, $j$, at the location, $\v{r}_b$, such that
\begin{align}
    \P = \sum\limits_j \P_j = \sum\limits_j \frac{k_{rj}^2|\v{k}_{hj}|^2}{\hat{\omega}_j K_j^2}\A_j \alpha_j.
\end{align}
\added[id=GSV]{Here, $\A_j = \N_j \Delta\kl \Delta\kp \Delta\kr$ and $\alpha_j$ are the ray volume's total wave action and the spatial fraction that the $j$th ray volume takes of the volume of the cell it is contained in, respectively.}
Thus, rather than each individual spectral component, we consider a spectrally integrated saturation. The reduction of the individual wave action components, $\A_j$, to their saturation value, $\A_j^*$, is then achieved through setting
\begin{align}
    \N_j \rightarrow \N_j^* = \N_j\left[1 - 2\kappa K_j^2 \Delta t\right],
\end{align}
with the turbulent diffusivity coefficient
\begin{align}
    \label{eq:turbulent-diffusivity}
    \kappa = \max\left(0,\frac{\P - \frac{\bar{\rho}}{2}}{2 \Delta t\sum\limits_j\P_j K_j^2}\right).
\end{align}
Saturation is thus triggered where the instability threshold \peqref{eq:instability-threshold} is violated by the constructive interference of locally overlapping internal waves. All wave amplitudes, or equivalently wave action densities, are then set to the corresponding saturation values. Note that \eqref{eq:turbulent-diffusivity} does not imply that the turbulent diffusivity diverges as $\Delta t \rightarrow\infty$. Because at the beginning of each time step $\P \le \frac{\bar{\rho}}{2}$, the numerator of the ratio is at best growing linearly with $\Delta t$, so that $\kappa$ always ends up finite. \added[id=GSV]{This spectral saturation implementation was shown to compare well to wave-resolving simulations in idealized cases by \cite{Boloni2016}. Furthermore, it is planned to couple the saturation scheme to the turbulence parameterization of the underlying atmospheric model, ICON. An in-depth analysis of the breaking processes and the impact of the spectral saturation on a global scale is hence left to a future study.}

\section{Atmospheric simulations with UA-ICON and MS-GWaM-3D}
\label{sec:results}

Building on \cite{Boloni2021} and \cite{Kim2021}, who introduced MS-GWaM-1D, we choose to work with the ICON model with its upper-atmosphere extension \citep{Zangl2015, Borchert2019b}. Here, we use version \added[id=GSV]{\textit{2.6.5-nwp1b}} with a horizontal resolution of approximately \qty{160}{\km} (model grid R2B04) and the physics packages for the numerical weather prediction (NWP) and the upper atmosphere. \added[id=GSV]{Note, that MS-GWaM replaces the non-orographic gravity wave drag but leaves the operational orographic wave parameterization intact \citep[in particular,][]{Lott1997}. Through the coupling to the mean wind, the latter thus indirectly interacts with MS-GWaM.} The setup has a model top of \qty{150}{\km} with vertical grid extent of a few tens of meters in the boundary layer, \qtyrange[range-units = single]{700}{1500}{\m} in the stratosphere, and a maximum of approximately \qty{4}{\km} in the lower thermosphere. A sponge layer acts above an altitude of \qty{110}{\km} which is why we restrict our analysis to altitudes below \qty{100}{\km}. The model is initialized with IFS analysis data below \qty{60}{\km} and with the climatological thermodynamic state at rest above. It is then spun up for a month to exclude adjustment effects in our analysis. These runs are repeated to simulate June and December for the years 1991 through 1998. Additionally, we deploy runs for June and December 1991 with extended diagnostics. In particular, the free parameters controlling the launch fluxes are the convective area fraction for the convectively generated waves and the summer and winter launch amplitudes of the background waves (see Sec. \ref{sec:raytracing}\ref{subsec:sources-sinks}). In general, stronger gravity-wave launch fluxes shift the wave saturation to lower altitudes, leading to a lower wind reversal at the mesopause. Comparisons of simulations with the horizontal wind model HWM \citep[hereafter HWM2014,][]{Drob2015} and the empirical temperature model NRLMSIS2.1 \citep[hereafter MSIS,][]{Emmert2022} then give a good estimate for the performance of the chosen parameters. \added[id=GSV]{The HWM2014 and MSIS are chosen based on their wealth of incorporated data in the middle and upper atmosphere in the time range of our simulations. In particular, they incorporate reanalysis, ground based datasets, and satellite based observations \citep[for more detail see][]{Drob2015, Emmert2021, Emmert2022}. The reference climatologies are then generated for exactly the simulated months.} With the help of short simulations spanning a large parameter space, we identified an optimum with an area fraction of $\varepsilon_q=\qty{7}{\%}$, a winter launch flux of $F_{w} = \qty{8}{\milli\Pa}$, and a summer launch flux of $F_{s} = \qty{2.4}{\milli\Pa}$ (for more detail, see Sec. \ref{sec:raytracing}\ref{subsec:sources-sinks}).

\subsection{Zonal mean flow}
\label{subsec:zonal-mean}

\begin{figure*}
    \centering
    \includegraphics[width=1\textwidth]{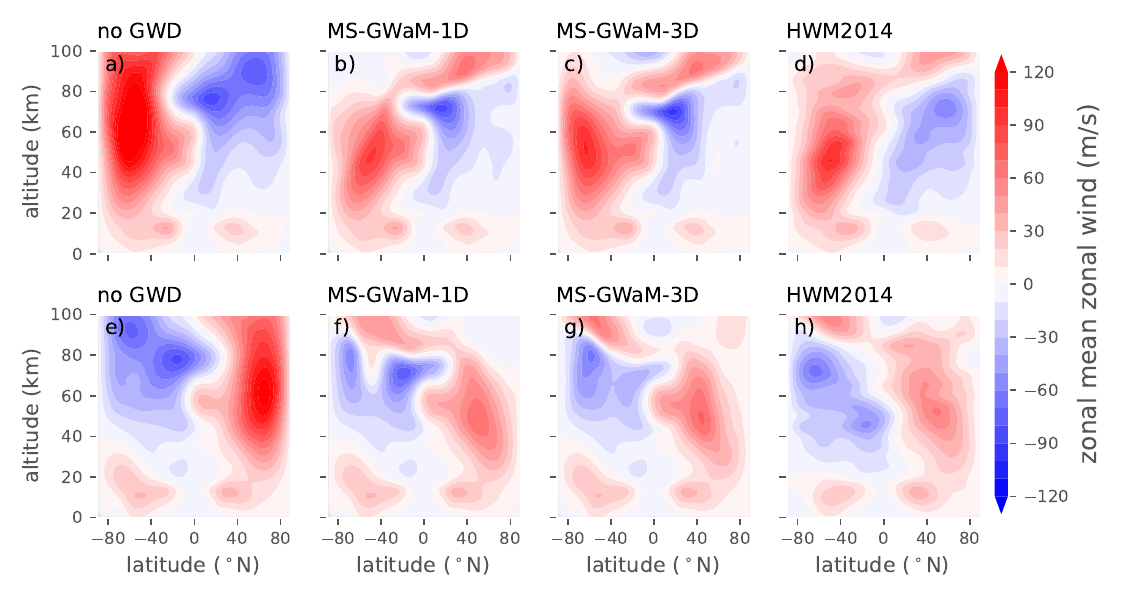}
    \caption{Zonal mean zonal winds from composites of runs for June (a-d) and December (e-h). The composites correspond to averages for the years 1991 through 1998. Runs from MS-GWaM-1D (b, f) and MS-GWaM-3D (c, g) are compared to simulations without any non-orographic IGW parameterization (a, e) and the HWM2014 climatologic model (d, h).}
    \label{fig:mean-zonal-wind}
\end{figure*}

\begin{figure*}
    \centering
    \includegraphics[width=1\textwidth]{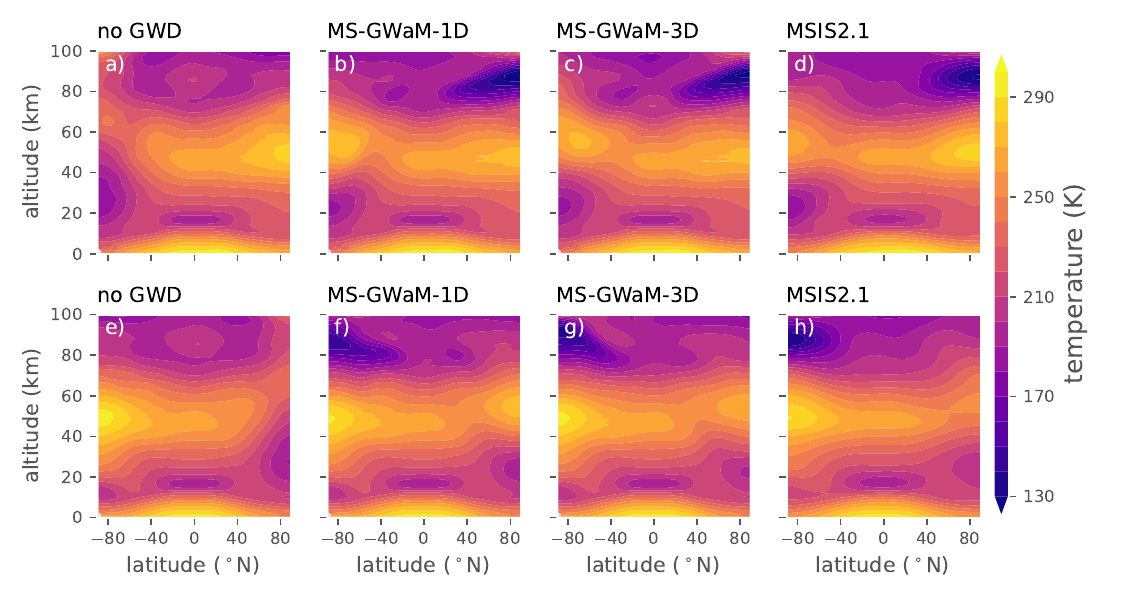}
    \caption{Like Fig. \ref{fig:mean-zonal-wind} but for zonally averaged temperatures. For comparison, we show the climatological model MSIS2.1 (d, h).}
    \label{fig:mean-temp}
\end{figure*}

To test the general performance of the gravity wave parametrization, we analyze the zonal-mean zonal wind and temperatures as composites over 8 simulations of June and December (Fig. \ref{fig:mean-zonal-wind} and \ref{fig:mean-temp}). To highlight the general effect of the parameterizations and visualize the general model performance, we accompany the data from simulations without non-orographic gravity wave parameterization. Naturally, the runs including a gravity wave drag parametrization mend model biases and thus differ significantly from the runs without. The comparison is, therefore, to be understood as a reference for the general behavior of ICON in the present setup and in how far model biases stem from the lack of a gravity wave drag parametrization or may be rooted elsewhere. Additionally, we show the reference climatologies HWM2014 and MSIS. Note that, as described above, both versions of MS-GWaM were tuned such that the mesopause height at $z\approx\qty{80}{\km}$ agrees with the HWM2014. In general, MS-GWaM-1D and 3D produce physical zonal-mean zonal winds with the expected wind reversals in both the summer and winter hemispheres (Fig. \ref{fig:mean-zonal-wind}b-d, f-h). As one may expect, runs with the same model setup but without any non-orographic parameterization are subject to strong jets and entirely lack the mesopause reversal (Fig. \ref{fig:mean-zonal-wind}a and e).

Notable differences to the HWM2014 can be seen in the summer hemispheric jets, in the tropical stratosphere, and in the Antarctic winter jet. Comparing the two MS-GWaM results with the run without any non-orographic gravity wave drag parameterization (Fig. \ref{fig:mean-zonal-wind}a and e) we find that the tropical stratospheric jet at $z\approx\qty{50}{km}$ height \added[id=GSV]{is a} robust feature of the current model version, which \added[id=GSV]{is} present in all three runs. The aforementioned tuning runs indicated that there is little to no influence on the structure \added[id=GSV]{by the tuning parameters} (not shown). We thus conclude that \added[id=GSV]{it is} generally independent of the gravity wave parameterization. \added[id=GSV]{The weak northern summer jet, however, seems to be related to the transient parameterization in general, as it is present in runs with both MS-GWaM-1D and MS-GWaM-3D (Fig. \ref{fig:mean-zonal-wind}b and c). While we note a similar bias in the work of \cite{Boloni2021}, the runs without any non-orographic gravity wave parameterization do not suffer from this bias (Fig. \ref{fig:mean-zonal-wind}d).} Finally, we observe that the slant of the Antarctic winter jet towards the equator is not entirely reproduced by MS-GWaM-3D (Fig. \ref{fig:mean-zonal-wind}c). This is due to \added[id=GSV]{a modified zonal and meridional gravity wave drag in the Southern hemisphere} above \qty{60}{km} altitude in MS-GWaM-3D as opposed to MS-GWaM-1D \added[id=GSV]{as will be discussed in Sec. \ref{sec:results}\ref{subsec:southern-jet-refraction}}. Possible reasons for \added[id=GSV]{the modified} wave drag might be improper spectral characteristics of the current background source, errors near the pole due to the representation of the ray-volume extent in spherical coordinate systems (c.f. Sec. \ref{sec:raytracing}\ref{subsec:repray}), or even co-dependencies of the columnar gravity wave parameterization with other parameterizations such as the \added[id=GSV]{the employed orographic wave drag parameterization. Moreover, other components of the model, such as the} radiation scheme or the modified ozone climatology, \added[id=GSV]{are tuned to produce realistic temperature distributions in combination with the standard orographic wave drag.} In general, the confidence in the present implementation is high, as MS-GWaM-1D was validated by \cite{Boloni2021, Kim2021} and MS-GWaM-3D was validated against the results of \cite{Wei2019} using an idealized cubic geometry implemented in ICON (not shown). However, a more detailed analysis may be needed to identify the reasons behind the phenomenon and mend the setup accordingly.

Similar to the zonally averaged zonal winds, we find that the zonally averaged temperatures and the corresponding MSIS fields agree well (Fig. \ref{fig:mean-temp}). In particular, the MS-GWaM reproduces the cold summer pole at the mesopause as well as the warm winter pole associated with the stratopause (Fig. \ref{fig:mean-temp}b-c and f-g). These features are naturally not present when the non-orographic gravity wave drag parameterization is switched off (Fig. \ref{fig:mean-temp}a and e).

\added[id=GSV]{In general,} the zonal-mean model state generally seems to be \added[id=GSV]{moderately} affected by enabling a 3-dimensional wave propagation in MS-GWaM. \added[id=GSV]{While the runs with the transient parameterizations generally excert some bias in the northern summer hemisphere, the Antarctic winter jet is too strong at altitudes around \qtyrange{50}{80}{\km} for MS-GWaM-3D. Runs for the month December do not contains these systematic biases.} This resembles the findings of \cite{Boloni2021} where a weak dependency of the zonal-mean wind on the transience of the gravity wave parameterization was detected. Albeit somewhat counter-intuitive, it may not be a surprise, as the spatial and temporal averages represent the quasi-steady state of the model setups. These in turn are tuned to approximate a similar zonal-mean model state. It was, however, also shown that both the transience and the localization of the gravity waves can have significant impacts on the structure of the horizontal mean flow \citep[e.g.][]{Sato2009, Sacha2016, Ehard2017}. For instance, the mean-state results may change if a more realistic spatio-temporal distribution of the wave source spectrum was employed. While the investigation of realistic source distribution is left for future work, in the following sections we focus on the nature of 3-dimensional propagation of parameterized IGWs, examining wave action fluxes and budgets as well as horizontal structures of IGW momentum fluxes and mean wave drags.

\subsection{Global 3-dimensional IGW distribution}

\begin{figure*}
    \centering
    \includegraphics[width=.9\textwidth]{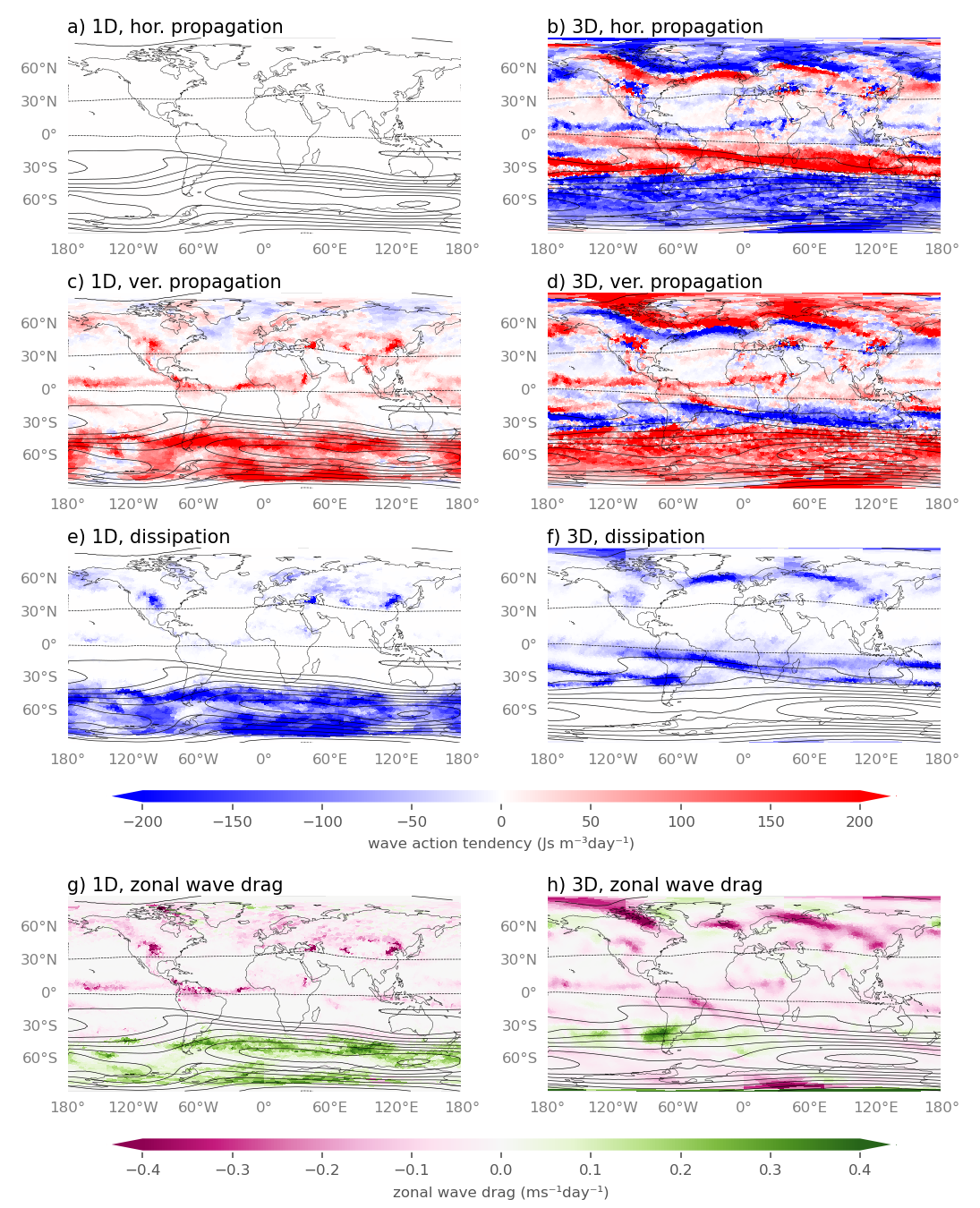}
    \caption{Wave action tendencies for MS-GWaM-1D (left column) and MS-GWaM-3D (right column) due to horizontal wave propagation (a, b), vertical wave propagation (c, d) and wave dissipation (e, f). The last row depicts the corresponding zonal wave drag. All tendencies are 10-day averages for June 10-20 1991 at an altitude of \added[id=GSV]{$z\approx\qty{25}{\km}$}. Additionally, \added[id=GSV]{contours of the zonal wind in steps of \qty{10}{\meter\per\second}} are shown.}
    \label{fig:wave-action-budget-25}
\end{figure*}

\begin{figure*}
    \centering
    \includegraphics[width=.9\textwidth]{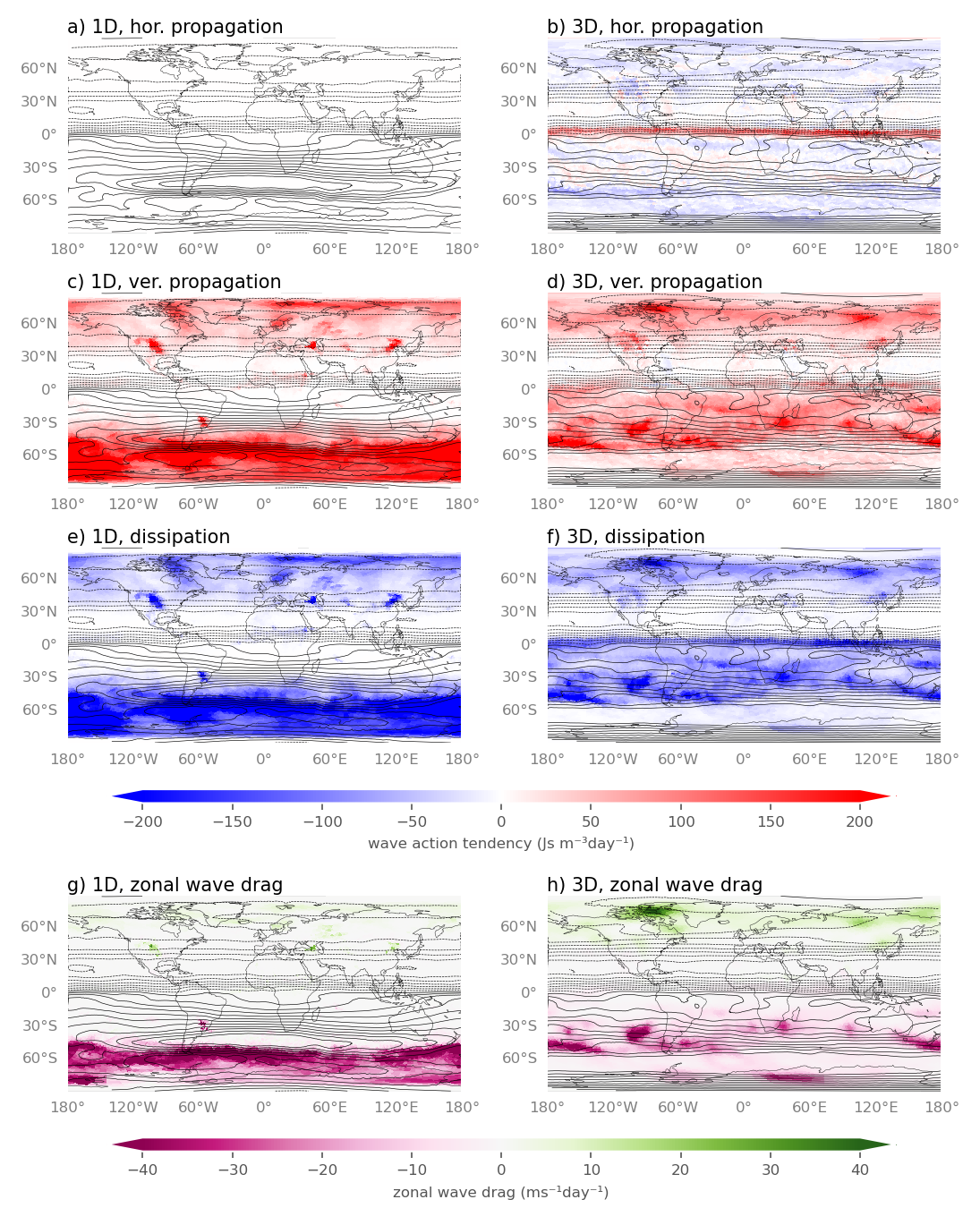}
    \caption{As Fig. \ref{fig:wave-action-budget-25} but for \added[id=GSV]{$z\approx\qty{60}{\km}$}}
    \label{fig:wave-action-budget-60}
\end{figure*}

Gravity waves are generally 3-dimensional phenomena that are present in large parts of the atmosphere. While wave-resolving simulations can \added[id=GSV]{provide} insights into global wave activity, they typically span short model times due to the limitations of both computational resources \added[id=GSV]{and} required storage. Additionally, the separation of IGWs from other dynamics and their characterization are generally non-trivial tasks. Being based on a multiscale WKBJ theory \citep{Achatz2017}, MS-GWaM simulates the global behavior of non-orographic gravity waves without the need to explicitly resolve their fast varying phases. As a result, it can be used to predict global distributions of phase-space wave action densities as well as derived quantities such as momentum fluxes and the resulting wave drag. As an example, we analyze the global wave action budget \added[id=GSV]{and augment it with the zonal wave drag to illustrate the resulting effect on the mean flow}.\\
Consider the phase-space integrated wave action conservation \peqref{eq:wa-6d-conservation}, averaged in time,
\begin{align}
    \label{eq:wave-action-budget}
    -\frac{\Delta \A}{\Delta t} &=
        \nabla_h\*\overline{\v{c}_{gh}\A}
        + \dr \overline{c_{gr}\A}
        + \overline{S},
\end{align}
where we define $(\v{c}_{gh}, c_{gr})\A = \int (\v{c}_{gh}, c_{gr})\N d^3k$, and the overbars denote the temporal average over the time interval, $\Delta t$. Here, the three right-hand side terms correspond to the temporally averaged wave-action tendency contributions from the horizontal wave propagation, the vertical wave propagation, and the wave saturation. Moreover, the left-hand term is small for sufficiently long averaging intervals. For a sufficiently small left-hand-side of \eqref{eq:wave-action-budget} while preserving the planetary wave structure, we choose $\Delta t = \qty{10}{\day}$ and show the predictions of the right-hand side contributions by MS-GWaM-1D and \added[id=GSV]{MS-GWaM}-3D at altitudes of approximately \added[id=GSV]{\qty{25}{\km} and \qty{60}{\km} for June 10-20, 1991. While the \qty{25}{\km} level shows the stark differences between the 1D and 3D schemes, the wave dynamics at \qty{60}{\km} altitude highlight the interaction with the Antarctic polar night jet in more detail.}

As one may expect, MS-GWaM-1D predicts a balance between the vertical \added[id=GSV]{wave action} propagation and dissipation \added[id=GSV]{at either altitude} (Figs. \ref{fig:wave-action-budget-25} and \ref{fig:wave-action-budget-60}, panels c and e). Thus, wave breaking and strong mean-flow \added[id=GSV]{impacts} can only occur where IGWs were previously able to vertically propagate through the wind shear of the underlying air column. \added[id=GSV]{Negative values for the change of wave action due to vertical propagation are associated with a finite temporal averaging interval, $\Delta t$. For instance, propagating mean-flow structures such as planetary waves may lead to changes in the wave action budget which adjust only on the time scale of the wave propagation. Although not linearly related, the strongest zonal wave drags occur where the wave dissipation is high. At the rather low altitude of \qty{25}{\km} this results in a rather uniform distribution of wave dissipation and drag in the Southern hemisphere, accompanied with intermittent convective events in the Northern hemisphere (Fig. \ref{fig:wave-action-budget-25}e to g)}. Running the same case with MS-GWaM-3D, we find that both the contributions due to the horizontal and vertical propagation are of similar magnitude \added[id=GSV]{at an altitude of \qty{25}{km}}, such that the resulting balance includes all terms (Fig. \ref{fig:wave-action-budget-25}b, d, and f). Near horizontally sheared wind structures, such as the edges of the Antarctic winter jet, the two contributions from the horizontal and vertical propagation form inverted dipoles associated with wave refraction phenomena. In the case of the mentioned Antarctic winter jet, as we will show below, this corresponds to the often observed wave refraction into the jet \citep[\eg][]{Ehard2017, Hindley2020}. Interestingly, the wave refraction follows the structure of the resolved planetary waves and ultimately leads to a shift of lower-level wave \added[id=GSV]{action} dissipation towards regions of strong horizontal shear, as for instance found between the summer and winter jets  (Fig. \ref{fig:wave-action-budget-25}f). Despite the different launch amplitudes (see Sec. \ref{sec:raytracing}\ref{subsec:sources-sinks}) all contributions have similar magnitudes on both hemispheres. \added[id=GSV]{Correspondingly, the horizontal propagation significantly modifies the zonal wave drag at \qty{25}{\km}. While the Southern hemispheric drag is much reduced, the Northern hemisphere exhibits comparatively strong drags near planetary wave structures. Interestingly, the wave action budgets change structurally with altitude. At $z\approx\qty{60}{\km}$, both the budgets of MS-GWaM-1D and MS-GWaM-3D are mostly controlled by vertical wave action propagation and dissipation (Fig. \ref{fig:wave-action-budget-60}c to f). In contrast to the lower altitude, the horizontal wave action propagation is much reduced. However, the horizontal propagation through lower altitudes shifts the action dissipation and wave drag near the Southern winter jet towards the equator (Fig. \ref{fig:wave-action-budget-60}e - h). We suspect this shift to cause the missing wave drag at higher altitudes near the pole, leading to a biased jet structure in the zonal mean. These results suggest an important influence of the horizontal propagation near the Antarctic winter jet and thus we analyze the the zonal mean drag structures in the following.}

\subsection{Wave propagation \added[id=GSV]{and drag} in the vicinity of horizontally sheared jet structures}
\label{subsec:southern-jet-refraction}

\begin{figure*}
    \centering
    \includegraphics[width=1\textwidth]{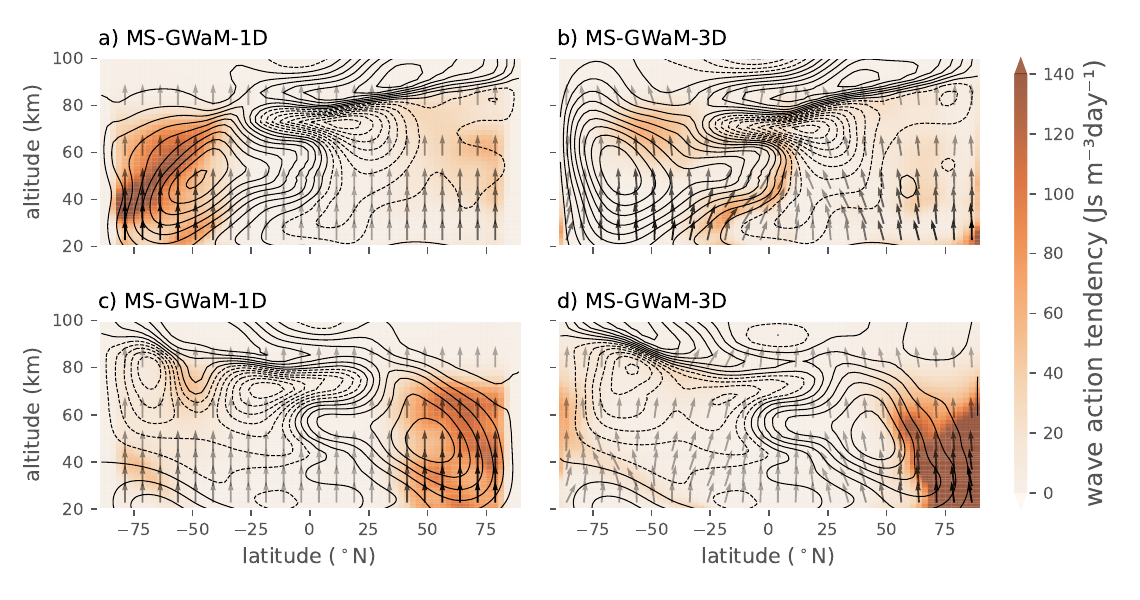}
    \caption{Wave action changes due to wave propagation in MS-GWaM-1D (a, c) and MS-GWaM-3D (b, d) for composites of June (a-b) and December (c-d). The field is overlaid with vectors of wave action fluxes. They are scaled with respect to the plot dimensions and units, the horizontal direction is enhanced by a factor of 3 for visualization. The arrow transparency represents the wave action flux strength. The contours represent the corresponding zonally averaged zonal wind with \qty{10}{\m\per\s} increments, with dashed and solid lines for negatives and non-negatives, respectively. See Fig. \ref{fig:gwd-global} for the corresponding wave drag.}
    \label{fig:wafl-global}
\end{figure*}
\begin{figure*}
    \centering
    \includegraphics[width=1\textwidth]{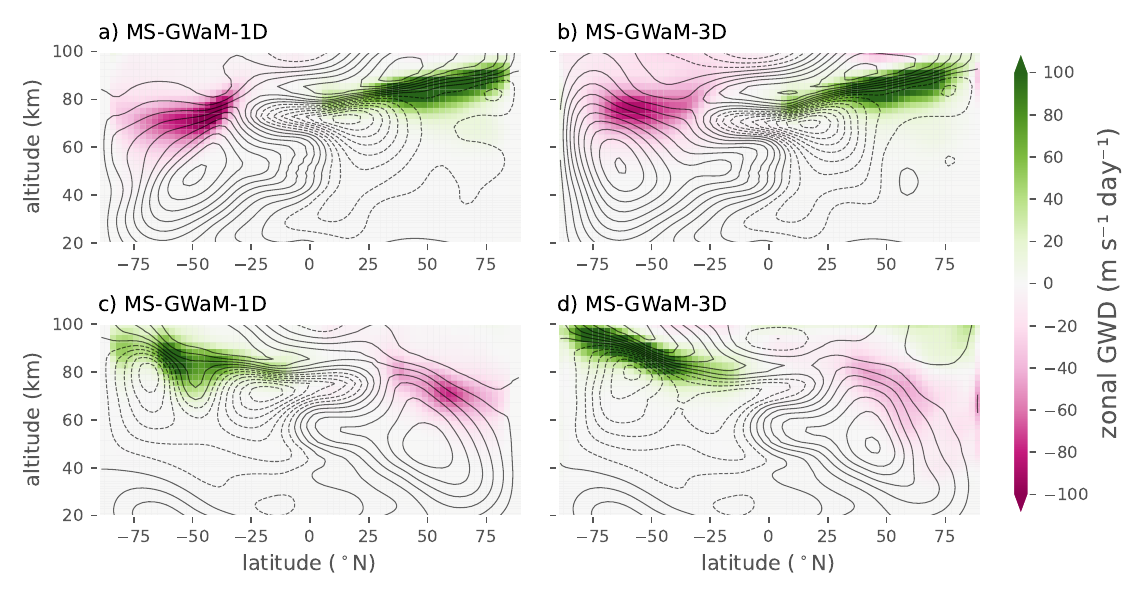}
    \caption{Zonally averaged zonal gravity wave drag due to the parametrized, non-orographic internal gravity waves. The results are averaged over composites of June (a-b) and December (c-d) 1991 through 1998 and plotted for both MS-GWaM-1D (a, c) and MS-GWaM-3D (b, d). The contours represent the corresponding zonally averaged zonal wind with \qty{10}{\m\per\s} increments.}
    \label{fig:gwd-global}
\end{figure*}
\begin{figure}
    \centering
    \includegraphics[width=.8\columnwidth]{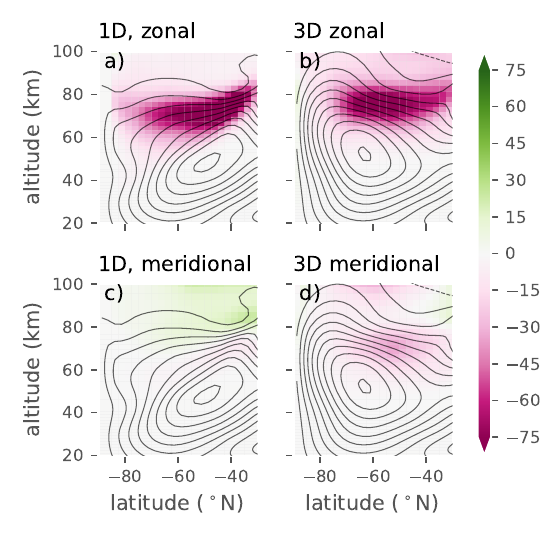}
    \caption{Zonally averaged zonal gravity wave drag due to the parametrized, non-orographic internal gravity waves. The results are averaged over composites of June (a-b) and December (c-d) 1991 through 1998 and plotted for both MS-GWaM-1D (a, c) and MS-GWaM-3D (b, d). The contours represent the corresponding zonally averaged zonal wind with \qty{10}{\m\per\s} increments.}
    \label{fig:gwd-jet}
\end{figure}

Expanding on the horizontal refraction behavior found in the horizontal maps (Figs. \ref{fig:wave-action-budget-25} and \ref{fig:wave-action-budget-60}) we show zonal averages of wave action fluxes (Fig. \ref{fig:wafl-global}) and the corresponding wave drag (Fig. \ref{fig:gwd-global}) temporally averaged over free runs \added[id=GSV]{for Junes and Decembers 1991 through 1998}. By construction, MS-GWaM-1D only allows for vertical propagation (Fig. \ref{fig:wafl-global}a and c). The corresponding wave-action tendency is thus associated with the vertical wave-action flux convergence. The consequently generated wave drag accelerates the zonal wind structures, with maxima occurring at the mesopause (Fig. \ref{fig:gwd-global}a and c). \added[id=GSV]{Note, that maximum wave action flux convergences are not equivalent to the maximum wave drag due to the scaling with the wave vector (compare Sec. \ref{sec:theory}).} MS-GWaM-3D shows strong horizontal propagation in regions of strong horizontal and vertical shear, such as jet edges (Fig. \ref{fig:wafl-global}b and d). Notably, in southern-hemispheric winter the wave action flux vector points northward near the South Pole, i.e. into the Southern winter jet, (Fig. \ref{fig:wafl-global}b). Consequently, the polar wave-action fluxes are reduced at altitudes near the mesopause and the wave drag maximum \added[id=GSV]{is weaker south of \qty{75}{\deg S}} (Fig. \ref{fig:wafl-global}b and Fig. \ref{fig:gwd-global}b). \added[id=GSV]{Moreover, the equatorial regions are strongly affected by the horizontal wave propagation (Fig. \ref{fig:gwd-global}b and d). Here, the wind shear is generally strong but the mean wave action fluxes are weak due to the intermittency of the convective wave source.\\}

\added[id=GSV]{To highlight the dynamics near the Antarctic winter jet, we additionally show the zonal and meridional wave drags for both MS-GWaM-1D and MS-GWaM-3D near the jet, averaged for all Junes (Fig. \ref{fig:gwd-jet}). Here, the reduced zonal wave drag for MS-GWaM-3D, south of \qty{75}{\deg S} becomes particularly visible (Fig. \ref{fig:gwd-jet}a - b). Moreover, the maximum zonal wave drag is shifted southward due to the horizontal propagation. Stronger differences can be seen for the generally weaker meridional component (Fig. \ref{fig:gwd-jet}c - d). In particular, it changes sign and exerts a southward drag on the jet when enabling the horizontal propagation. Note that this behavior is only observed above the jet. Globally, the meridional drag accelerates the flow from the winter towards the summer hemisphere for both models (not shown). It remains unclear, however, whether the modeled meridional drag is a consequence of the too strong winter jet over the pole or vice versa. For instance, the ray volume discretization, the interaction with the orographic wave drag parameterization or the structure of the modeled temperatures may be a cause of the biased wind structures (also compare Sec. \ref{sec:results}\ref{subsec:zonal-mean} and Fig. \ref{fig:mean-zonal-wind}c).} Further and more targeted investigation may be needed to bring clarity about this effect.

\added[id=GSV]{Differences in} refraction behavior can also be seen in the summer hemispheres. However, the model biases of the Northern summer hemispheric jet structures (c.f. Sec. \ref{sec:results}\ref{subsec:zonal-mean}) complicate the interpretation of the wave action budgets in the corresponding regions.\\

\begin{figure}
    \centering
    \includegraphics[width=.8\columnwidth]{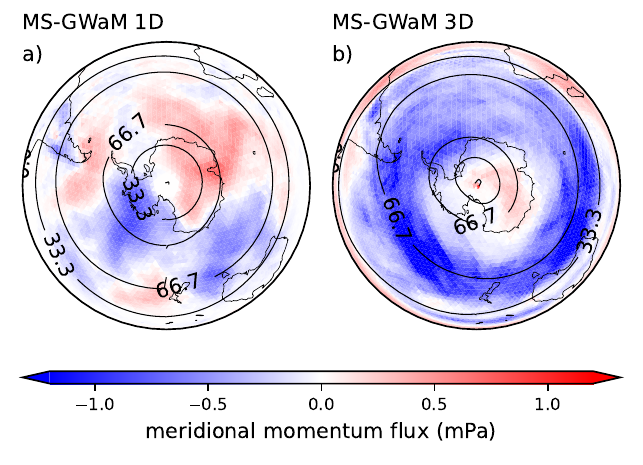}
    \caption{Meridional momentum fluxes in MS-GWaM-1D (a) and MS-GWaM-3D (b) for composites of Junes from 1991 through 1998 at an altitude of \added[id=GSV]{$z\approx\qty{40}{\km}$} overlaid by the corresponding zonal winds (contours, units in \qty{}{\m\per\s}). The maps show the typically observed refraction of internal gravity waves into the Southern polar night jet and modified wind patterns.}
    \label{fig:meridional-mflx}
\end{figure}

\added[id=GSV]{One may now ask the question, how realistic the modeled meridional wave propagation near the Antarctic winter jet is and how it corresponds to satellite observations. To provide some insight,} we show the meridional momentum fluxes (\cf Eq. \ref{eq:fluxes:momentum}) from the composites of June for the years 1991 through 1998 \added[id=GSV]{at altitudes $\approx\qty{40}{\km}$} (Fig. \ref{fig:meridional-mflx}). Note that the meridional fluxes of MS-GWaM-1D must be interpreted keeping in mind the assumed local horizontal homogeneity. That is, albeit the wave parameters suggest horizontal momentum fluxes, the horizontal propagation (including the wave advection) are set to zero in the columnar setup. The resulting \textit{synthetic} meridional fluxes seem to be unrelated to the underlying \added[id=GSV]{zonal} jet structure (Fig. \ref{fig:meridional-mflx}a). As waves are launched homogeneously in all directions at the lower boundary (Sec. \ref{sec:raytracing}\ref{subsec:sources-sinks}) the resulting meridional fluxes are a product of the partial wave filtering due to local vertical gradients of both the horizontal wind and the buoyancy frequency. In contrast, MS-GWaM-3D shows meridional convergence of wave momentum fluxes into the polar jet (Fig. \ref{fig:meridional-mflx}b). \added[id=GSV]{In particular, the meridional fluxes are northward over the Antarctic continent and southward further north. The change in sign, i.e. the convergence zone, roughly coincides with the zonal wind maximum. This convergence effect of internal gravity waves into the polar night jets is well documented from both observations and general circulation models \citep[e.g.][]{Dunkerton1984, Sato2012, Ehard2017, Hindley2020, Gupta2021}. Notably, the structure and amplitudes of the meridional momentum fluxes from MS-GWaM-3D simulations resemble the observations of \citeauthor{Hindley2020} (\citeyear{Hindley2020}, compare their Fig. 3d). Naturally, MS-GWaM cannot reproduce the strong momentum fluxes originating from orographic waves over the Antarctic peninsula, the Southern Andes or the islands of the Southern Ocean. The otherwise fairly close fit, however, gives us high confidece in the correctness of MS-GWaM-3D.}

\section{Discussion of achievements and challenges}
\label{sec:conclusions}

Although internal gravity waves (IGWs) are an essential part of atmospheric dynamics, their parameterization in general circulation models is typically subject to various simplifications. In particular, the single-column assumption, the steady-state approximation, and the balanced mean-flow assumption have usually been made. Here, we present a fully 3-dimensional implementation of the Multi-Scale Gravity Wave Model, MS-GWaM-3D, which aims at parameterizing IGWs through Lagrangian ray-tracing without the three mentioned assumptions. In general, we find that including 3D transience significantly impacts global IGW propagation patterns and thus horizontal IGW distributions. Correspondingly, the associated wave drag and mean flows are modified.

We compare two distinct situations: For MS-GWaM-1D, where the columnar approximation is applied, the vertical propagation is balanced by wave breaking in the wave action equation (Figs. \ref{fig:wave-action-budget-25} and \ref{fig:wave-action-budget-60}, panels c and e). Only waves that did not encounter critical layers according to their wave properties may propagate to higher altitudes and eventually break. In contrast, MS-GWaM-3D allows for modulated and spatially unconstrained propagation and enables waves to be refracted around wind structures. As a consequence, both the horizontal and vertical propagation balance with the wave dissipation (Figs. \ref{fig:wave-action-budget-25} and \ref{fig:wave-action-budget-60}, panels b, d, and f). The equal order of magnitude in the contributions of the horizontal and vertical wave propagation to the wave action budget emphasizes the importance of including horizontal propagation. Moreover, it questions the validity of using columnar methods when, e.g., studying IGW distributions. Both methods do, however, perform as expected in reproducing the cold summer pole and the warm winter pole at altitudes \qty{\sim85}{\km} and \qty{\sim60}{\km}, respectively, in the climatological zonal mean (Fig. \ref{fig:mean-temp}). The corresponding wind reversals and the mesopause altitudes are reasonably predicted (Fig. \ref{fig:mean-zonal-wind}). We are thus confident that MS-GWaM-3D, as MS-GWaM-1D before it, covers some major effects of IGWs on the mean-flow dynamics, rendering MS-GWaM-3D a viable IGW parameterization. The 3D wave propagation is also found to have significant impacts on the variability of the zonal-mean flow, such as the quasi-biennial oscillation (QBO), \added[id=GSV]{as presented by \cite{Kim2023a}}.

Including the horizontal propagation does, however, also introduce some important differences in the simulated climatology. In particular, the southern hemispheric winter jet becomes stronger near the pole at altitudes \qtyrange[range-units = single]{60}{80}{\km} for MS-GWaM-3D (Fig. \ref{fig:mean-zonal-wind}c). The reason behind this change is, however, not obvious. One possible explanation is the northward refraction of \added[id=GSV]{gravity waves} near the Antarctic winter jet in MS-GWaM-3D as compared to MS-GWaM-1D. Both the 3D wave action tendencies (dipole structures in Fig. \ref{fig:wave-action-budget-25}d and f) and the zonally averaged wave action fluxes (vectors in Fig. \ref{fig:wafl-global}a and b) suggest that the 3D modulation causes IGWs to propagate northward and thus relate to a weaker gravity wave drag at high altitudes and similar latitudes (Fig. \ref{fig:gwd-jet}a and b). \added[id=GSV]{Additionally, the meridional wave drag near the jet is oppositely directed, that is southward, for MS-GWaM-3D (Fig. \ref{fig:gwd-jet}c and d). It remains unclear, however, whether the wave refraction and ultimately the wave drag cause the shifts of the jets as the two structures a non-linearly coupled. Possible other causes are the numerical simplifications made in the setup of the ray tracer in the spherical coordinate system or codependencies of the gravity wave parameterization with other parameterizations, for instance the orographic gravity wave drag. Finally, the jets are sensitive to the structure of the modeled temperature distribution, which in turn is dependent on for instance the ozone climatology and the radiation scheme. A thorough investigation will be needed, and we hope to answer these questions in an envisioned follow-up study. Finally, we find that the meridional wave propagation in MS-GWaM-3D around the Antarctic polar night jet resembles the observations of \cite{Hindley2020} both in amplitudes and structure, up to orographic wave sources such as the Southern Andes (Fig. \ref{fig:meridional-mflx}). This gives us high confidence into the implementations as the convergence into the polar night jet is well documented in a number of studies \citep[e.g.][]{Dunkerton1984, Sato2012, Ehard2017, Hindley2020, Gupta2021}.}

There is a long list of effects that may be analyzed in more detail from this starting point. Notably, the role of IGWs in sudden stratospheric warmings \citep[e.g.][]{Baldwin2021}, the accuracy of the final warming date \citep[e.g.][]{Camara2016, Eichinger2023}, the missing wave drag at \qty{60}{\deg S} \citep[e.g.][]{Holt2023}, etc. are of interest and are planned for future simulations and analyses. Current investigations include the impact of 3D-transient waves on the quasi-biennial oscillation and wave intermittency. Moreover, there are efforts into increasing the realism of the included IGW sources related to jets-frontal systems and flow over topography. Finally, code optimization may reduce the computational cost which amounts to a runtime factor $\sim30$ with respect to runs without any non-orographic wave drag parameterization (c.f. Appendix \ref{app:performance}). With these challenges in mind, we ultimately aim for the application of MS-GWaM-3D in climate simulations and possibly numerical weather predictions. 

Furthermore, there are numerous potential improvements that may be added to MS-GWaM in the future, albeit not considered here. Most importantly, MS-GWaM-3D---for the time being---does not include a flow-dependent description of the emission of IGWs from jets and fronts \citep{Charron2002, Richter2010, Camara2015} or flow over topography \citep{Palmer1986, Bacmeister1994, Lott1997, Xie2020, Xie2021, vanNiekerk2021, vanNiekerk2023, Eichinger2023}. For the latter reason, MS-GWaM needs augmentation with a subgrid-scale orography parameterization. Moreover, it is not clear how important modulated triadic resonant interactions among IGWs or between subgrid-scale geostrophic modes and IGWs are for atmospheric dynamics, which could, however, be investigated through ray tracing techniques \citep[c.f.][]{Kafiabad2019, Voelker2020}.

%

\acknowledgments

UA thanks the German Research Foundation (DFG) for partial support  through the research unit "Multiscale Dynamics of Gravity Waves" (MS-GWaves, grants Grants AC 71/8-2, AC 71/9-2, and AC 71/12-2) and CRC 301 "TPChange" (Project-ID 428312742, Projects B06 “Impact of small-scale dynamics on UTLS transport and mixing” and B07 “Impact of cirrus clouds on tropopause structure”). 
YHK and UA thank the German Federal Ministry of Education and Research (BMBF) for partial support through the program Role of the Middle Atmosphere in Climate (ROMIC II: QUBICC) and through grant 01LG1905B.
UA and GSV thank the  German Research Foundation (DFG) for partial support  through the CRC 181 “Energy transfers in Atmosphere and Ocean” (Project Number 274762653, Projects W01  “Gravity-wave parameterization for the atmosphere” and S02 “Improved Parameterizations and Numerics in Climate Models.”). UA is furthermore grateful for support by Eric and Wendy Schmidt through the Schmidt Futures VESRI “DataWave” project. This work used resources of the Deutsches Klimarechenzentrum (DKRZ) granted by its Scientific Steering Committee (WLA) under project ID bb1097.

 %

\datastatement

All data used in this publication may be available upon request to the corresponding author.


\appendix[A] 

\appendixtitle{The gravity wave forcing of the mean flow and its relation to the EP-flux formulation}
\label{app:ep-flux-relation}

\added[id=GSV]{As laid out in Sec. \ref{sec:theory}, MS-GWaM-3D utilizes the perturbation flux formulation as introduced by \cite{Wei2019}. The interested reader may note that the resulting mean flow impact differs from the commonly used pseudomomentum flux convergence related to the Eliassen-Palm flux \citep{Eliassen1961, Andrews1976}. To highlight the relation to the latter, we would first like to remind the reader about the underlying assumption of the multiscale theory applied here. Firstly, the theory is built on the compressible Euler equations such that non-linearities not only arise from the advection operators but also from the pressure gradient terms. As MS-GWaM acts as a gravity wave parameterization for the underlying Eulerian model, phase averages are constructed in a Eulerian sense, accordingly. In terms of scale separation, it is assumed that horizontal scales are much larger than vertical scales. Furthermore, the Brunt-Väisälä frequency is assumed to be much larger than the Coriolis frequency. These assumptions then lead to the regime of small Rossby numbers with a leading order mean-flow in hydrostatic and geostrophic balance. Finally, all wave scales are assumed to be much smaller than the synoptic mean-flow scales.\\
Under these conditions, the impact of the internal gravity waves on the mean momentum comprises the momentum-flux convergence and an elastic term arising from variations in density and pressure in the compressible atmosphere. In addition, the mean entropy is forced by a wave potential temperature flux convergence. More specifically, the mean flow impact can then be written as}
\begin{align}
    \label{eq:mean-flow-impact-u-revisited}
    \left.\dt \v{U}\right|_{IGW} =
        &-\frac{1}{\bar{\rho}}\nabla\*(\bar{\rho}\left\langle\v{v}'\v{u}'\right\rangle)\\
        &+ \frac{f}{g}\v{e}_r\times\left\langle \v{u}'b'\right\rangle,\nn\\
    \label{eq:mean-flow-impact-t-revisited}
    \left.\dt \theta\right|_{IGW} = 
        &-\nabla\*\left\langle \v{u}'\theta'\right\rangle.
\end{align}
\added[id=GSV]{with the wave fluxes expressed as}
\begin{align}
    \label{eq:fluxes-momentum-revisited}
    \bar{\rho}\left\langle \v{v}'\v{u}'\right\rangle &= \int \left[\frac{
        \omh^2\cgh\vkh\N + f^2(\v{e}_r\times\cgh)(\v{e}_r\times\vkh\N)
        }{\omh^2 - f^2}\right]d^3k,\\
    \label{eq:fluxes:temperature-revisited}
    \left\langle \v{u}'\theta'\right\rangle
    &= \int \left[
        \frac{\bar{\theta}}{g\bar{\rho}} \hat{c}_{gr} \frac{fN^2}{\omh^2 - f^2} \v{e}_r\times\vkh\N
        \right]d^3k,\\
    \label{eq:fluxes:buoyancy-revisited}
    \frac{f}{g}\er\times\left\langle \v{u}'b'\right\rangle 
    &= \frac{f}{\bar{\theta}}\er\times\left\langle \v{u}'\theta'\right\rangle.
\end{align}
\added[id=GSV]{It is now interesting to additionally consider the \added[id=UA]{quasigeostrophic} potential vorticity evolution in the presence of internal gravity waves. The corresponding prognostic equation reads}
\begin{align}
    \label{eq:qgpv}
    \left(\dt + \v{U} \* \nabla_{\v{r}}\right)P &= -\v{e}_r\*\nabla_{\vr}\times\left[\frac{1}{\rhobar}
    \begin{pmatrix}
        \nabla_{\vr}\* \H \\
        \nabla_{\vr}\* \G \\
        0
    \end{pmatrix}
    \right],
\end{align}
\added[id=GSV]{where $P = \nabla_{\vr}^2\psi + \bar{\rho}^{-1}\dz(\bar{\rho}f^2/N^2\dz\psi)$ is the quasigeostrophic potential vorticity, denoted in terms of the mean-flow stream function, $\psi$. Note, that the latter relates to the leading-order synoptic-scale wind through $\v{U} = \v{e}_r\times\nabla_{\v{r}}\psi$ \added[id=UA]{and that within geostrophic theory also the thermodynamic fields can be obtained to leading order from the streamfunction. Hence, as long as the mean flow is within the synoptic scaling regime, it is controlled completely by \eqref{eq:qgpv}}. Furthermore, we have denoted the horizontal pseudomomentum fluxes, $\H$ and $\G$, defined by the spectral integral over the phase-space pseudomomentum fluxes, $\G = \int k\vcgh\N d^3k$ and $\H = \int l\vcgh\N d^3k$. For a full derivation, we refer the reader to the work of \cite{Achatz2017} and \cite{Achatz2022}, a recent summary was given by \cite{Achatz2023b}.} \added[id=UA]{Of interest is now that, \emph{if the mean-flow is in geostrophic and hydrostatic balance}, \eqref{eq:qgpv} can also be derived from a modified system of mean-flow equations where one replaces}
\begin{align}
    \label{eq:fluxes-momentum-ep-flux}
    \bar{\rho}\left\langle \v{v}'\v{u}'\right\rangle &\rightarrow \int \cgh\vkh\N d^3k,\\
    \label{eq:fluxes:temperature-ep-flux}
    \left\langle \v{u}'b'\right\rangle 
    &= \frac{g}{\bar{\theta}}\left\langle \v{u}'\theta'\right\rangle \rightarrow 0,
\end{align}
\added[id=UA]{and deletes the elastic term. This is the simplification that is most commonly applied in gravity-wave parameterizations: The momentum-flux convergence in the momentum equation is replaced by the convergence of pseudomomentum flux, the elastic term is neglected, and the entropy flux is set to zero. Under the conditions stated the mean flow then follows to leading order the same dynamics as predicted by (\ref{eq:mean-flow-impact-u-revisited}) and (\ref{eq:mean-flow-impact-t-revisited})}\\

\added[id=UA]{Since the zonal means of zonal wind and temperature are to a good approximation in geostrophic and hydrostatic balance, we therefore do not find a statistically significant difference in those fields.} However, \cite{Wei2019} showed that the \added[id=UA]{local} transient response of the Eulerian phase averaged wind in the presence of a finite amplitude gravity wave packet becomes more accurate when considering the fluxes as denoted in \eqrefs{eq:mean-flow-impact-u-revisited}{eq:mean-flow-impact-t-revisited}. We thus implement this formulation but leave the in-depth analysis of the local wind response to an envisioned follow-up study.

\appendix[B] 

\appendixtitle{Details of the numerical implementation}

To simulate transient internal gravity waves, MS-GWaM uses a spectral extension of WKBJ theory \citep[and references therein]{Bretherton1966, Grimshaw1975, Achatz2017, Achatz2022}. A Lagrangian approach is employed to solve the equations numerically \citep{Muraschko2015, Boloni2016, Boloni2021, Kim2021}. Therein the 6-dimensional phase space, spanned by position and wave number, is split into small finite-size ray volumes, propagating according to the eikonal equations \peqrefr{eq:group-velocity}{eq:eikonal-sphere:3}. They keep their regular shape of rectangles in $(\lambda,\phi,r,\kl,\kp,\kr)$ but their extent in all these six directions are allowed to vary in a volume-preserving manner. The wave-action density carried by each ray volume is constant, unless wave breaking leads to the onset of turbulence and hence a decrease in $\N$. These discrete Lagrangian volumes must, however, be connected to the local Eulerian model state to incorporate direct non-linear interactions between the mean flow and the gravity waves (c.f. Sec. \ref{sec:theory} and \ref{sec:raytracing}). This coupling poses a number of computational challenges, which are solved as described below.

\begin{figure}
    \centering
    \includegraphics[width=.8\columnwidth]{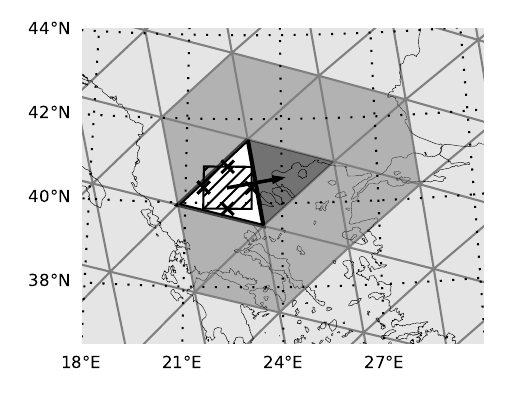}
    \caption{Sketch of the horizontal extent of a Lagrangian ray volume propagating in the unstructured triangular grid of the ICON model. The propagating ray volume (hatched box) may be associated with a current parent column (framed white grid cell). Given a horizontal group velocity (black arrow) we may expect it to propagate into a neighboring column (dark gray shaded grid cell). The crosses at the edges of the ray volume mark the locations for which background field data is needed to calculate the propagation of the ray volume.}
    \label{fig:horizontal_geometry}
\end{figure}

\section{3-dimensional field interpolation}
\label{app:interpolation}

When calculating the propagation of a finite-size ray volume in an inhomogeneous medium, it is important to consider that even under the assumption that all wave numbers are constant throughout the volume, the group velocities can differ significantly within the same ray volume. In particular, the scale height correction, $\Gamma$, the Brunt-Väisälä frequency, $N$, and the mean winds, $\v{U}$, are functions of space and differ between the four faces forming with the meridional and zonal boundaries of the ray volume (see crosses in Fig. \ref{fig:horizontal_geometry}). To account for a corresponding change in the area of the ray volume, we interpolate all needed background fields horizontally to the face-center positions and the ray volume center through linear Taylor expansions as follows:
\begin{align}
    \chi(\v{r}\sub{j}) = \chi(\v{r}\sub{cc}) + \left[\nabla_{\v{r}}\chi\right]\sub{cc}\*(\v{r}\sub{j} - \v{r}\sub{cc}),
\end{align}
where $\chi$ symbolizes the interpolated field and the subscripts \textit{cc} and \textit{j} refer to evaluations at the cell center and the ray-volume face or ray-volume center, respectively. In the vertical, we apply linear interpolations between the grid points. While the tendencies for the wave numbers are computed with the ray-volume centered values, the interpolated fields at the ray volume faces are used to compute group velocities, enabling the prediction of the changing extent of the ray volume in physical space \peqrefr{eq:dispersion}{eq:eikonal}. In particular, the difference in the group velocities at the center of opposing ray-volume faces predicts the rate of compression along the corresponding direction. The average of the opposing group velocities is then used to predict the change of location along the characteristic \peqref{eq:group-velocity}. Consequently, the change in the wave number extents $\Delta(\kl,\kp,\kr)$ are determined following the procedure described in section \ref{sec:raytracing}\ref{subsec:repray}. 

\section{Horizontal propagation and code parallelization}
\label{app:horizontal_propagation}

Since MS-GWaM propagates Lagrangian ray volumes through space, it cannot easily be parallelized using the traditional MPI infrastructure of the ICON model developed for the synchronization of Eulerian fields. In particular, the ray volumes may propagate over large distances and transfer between parallel MPI domains. To accompany this while avoiding load balancing problems, we associate each ray volume with a parent cell based on its current location and store it in a corresponding grid-based array. As an example, in the shown sketch (Fig. \ref{fig:horizontal_geometry}) the hatched ray volume is associated with the white-bounded underlying cell. It may, however, have a group velocity such that it propagates into a new parent cell, represented by the arrow and the dark gray shaded triangle, respectively. Therefore, after each propagation step, all cells such as the dark gray shaded triangle search all rays in all neighboring cells (light gray shading) and transfer those that are now located in their own cell area. With this construction, we can localize arrays of ray volumes based on the Eulerian grid column. The integration along the wave characteristics, launch processes, saturation calculations, and wave projection onto the Eulerian grid can then be parallelized using the domain decomposition of the dynamical core with the given model infrastructure. Note that the modulation equations are integrated using a 3-step Runge-Kutta low-storage scheme \citep{Williamson1980}.

\section{Ray volume splitting}
\label{app:splitting}

In general, the numerical treatment of internal gravity waves as ray volumes is set up to be exact in the limit of infinitesimally small ray volumes and Eulerian grid cells. The actual sizes of both are, however, finite. While the total phase-space volume of the rays is constant by construction, they may stretch or compress either in physical or spectral space. To avoid large numerical artifacts, we therefore split all ray volumes, $j$, vertically which become larger than the local thresholds
\begin{align}
    \Delta r_j > \max (\alpha_s h_c, \qty{1}{\km}),
\end{align}
with the local grid-cell height, $h_c$, and a scale factor $\alpha_s$ (which is set to 2.5 in this study). The volume is cut into two equally sized ray volumes in the vertical, with the wave vectors identical to the original parent wave. The carrier ray location of each split volume is then set to its center. Similarly, ray volumes are split horizontally where either of the three thresholds is reached
\begin{align}
    \cos\phi_j\Delta\lambda_j & > \alpha_s \sqrt{a_c} / R,\nn\\
    \Delta\phi_j & > \alpha_s \sqrt{a_c} / R,\\
    \cos\phi_j\Delta\lambda_j\Delta\phi_j & > \alpha_s a_c / R^2.\nn\\
\end{align}
Here, $a_c$ represents the local cell area, and $R$ the Earth mean radius. Splitting is done along the direction of the larger extent.

\section{Horizontal flux smoothing}
\label{app:smoothing}

To avoid numerical artifacts based on the employed simplification in the ray volume representation, we smooth all diagnostic outputs and wave fluxes. Each triangular cell combines three vertices, one in each corner (c.f. Fig. \ref{fig:horizontal_geometry}). Each of these vertices is connected to five or six neighboring cells, depending on the location of the triangulated icosahedron. The smoothing algorithm first averages the corresponding fields over these pentagons or hexagons and then combines the three resulting values into one average. Consequently, the weights are distributed such that the center cell is weighted with a factor $3$, the direct adjacent cells with one common edge are weighted with a factor $2$, and all neighboring cells sharing a vertex with the center cell are weighted with a factor $1$. While this procedure generally reduces the amplitude of strong flux peaks, it also reduces the occurrence of grid-scale noise and ensures model stability.

\section{Computational performance}
\label{app:performance}

\begin{figure}
    \centering
    \includegraphics[width=.8\columnwidth]{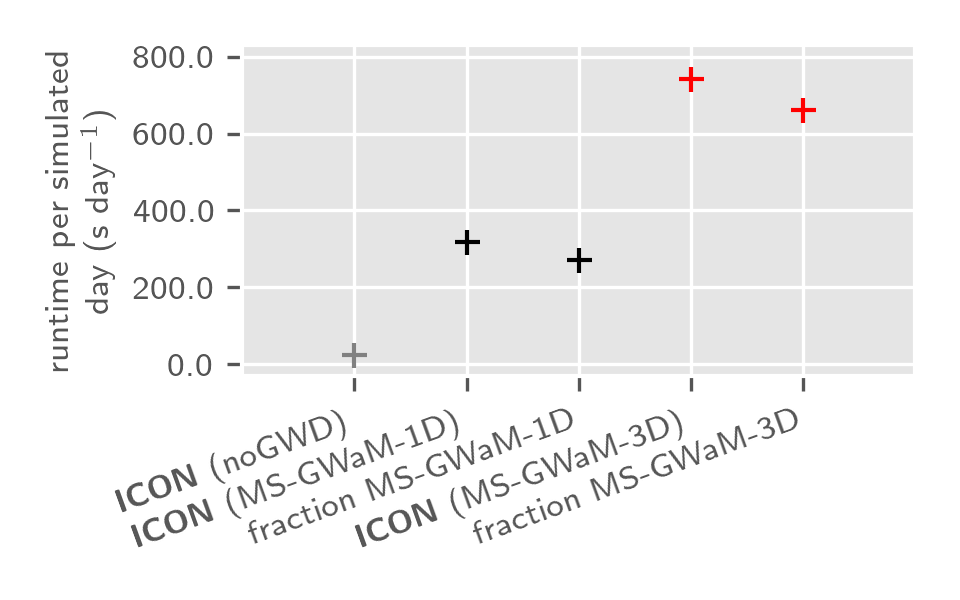}
    \caption{Performance comparison for UA-ICON at the R2B4 resolution (\qty{\sim160}{\km}) parallelized over 1920 MPI cores for runs without IGW parameterization (gray), MS-GWaM-1D (black) and MS-GWaM-3D (red). In addition to the total run time of ICON per simulated day, we show the fraction of run time associated with MS-GWaM-1D and -3D. The median values for the three setups are \qty{25}{\s\per\day}, \qty{317}{\s\per\day}, and \qty{743}{\s\per\day} and thus MS-GWaM-1D and-3D slow down the model by about a factor of 12 and 30, respectively.}
    \label{fig:performance}
\end{figure}

Finally, conceptional models need to prove themselves not only through the accurate representation of not-resolved physics but also through competitive performance. In particular, ray-tracing models may suffer from load balancing problems due to the uneven distribution of ray integrations between processors but also the existence of too many rays in general. As described in Sec. \ref{app:horizontal_propagation}, MS-GWaM utilizes a cell-based representation of ray volumes which is highly parallelizable such that load balancing problems may not inhibit the model performance. Additionally, the total number of ray volumes is restricted by the maximum numbers $n_{bg}\leq 4800$ and $n_{conv} \leq 8800$ per model-grid column for background and convective gravity waves, respectively. For the lowest impact, the ray volumes with the lowest wave energy are chosen to be removed. Detailed analyses show that the impact of this procedure is indeed small and may be neglected (not shown).

To measure the performance of the model, we compare the runtimes of all two-month runs without any non-orographic gravity wave parameterization with both the MS-GWaM-1D and MS-GWaM-3D setups (Fig. \ref{fig:performance}). Additionally, we show the fraction of runtime that is associated with MS-GWaM and normalize all values with the number of simulated days. These measures show that MS-GWaM-1D and MS-GWaM-3D slow down ICON by factors of approximately 12 and 30, respectively. That is, the 3D setup has a performance penalty equal to a factor of approximately 2.5 Still, the gravity wave parameterization is several orders of magnitude faster than wave-resolving simulations - for which \cite{Boloni2021} estimate a cost factor of about six orders of magnitude - and thus a competitive tool for the estimate and analysis of global IGW effects.


\bibliographystyle{ametsocV6}
\bibliography{refs-abbrevs,refs_zotero}

\end{document}